\documentclass[nofootinbib,twocolumn,showpacs,amsmath,amstex,amssymb,mathfonts,prb]{revtex4-2}

\usepackage{graphicx}
\usepackage{color}
\usepackage{bm}
\usepackage{amsmath}
\usepackage{amssymb}
\usepackage{amsthm}
\usepackage{amsfonts}
\usepackage[colorlinks,bookmarks=true,citecolor=blue,linkcolor=red,urlcolor=blue]{hyperref}

\usepackage{bbm}

\newcommand*{\id}{{\normalfont\hbox{1\kern-0.15em \vrule width .8pt depth-.5pt}}}

\newcommand{\be}{\begin{equation}}
\newcommand{\ee}{\end{equation}}
\newcommand{\bq}{\begin{eqnarray}}
\newcommand{\eq}{\end{eqnarray}}

\theoremstyle{theorem}

\theoremstyle{theorem}

\theoremstyle{definition}

\theoremstyle{definition}

\theoremstyle{remark}

\theoremstyle{theorem}

\makeatletter
\def\@fnsymbol#1{\ensuremath{\ifcase#1\or \dagger\or \ddagger\or
   \mathsection\or \mathparagraph\or \|\or **\or \dagger\dagger
   \or \ddagger\ddagger \else\@ctrerr\fi}}
    \makeatother

\begin{document}

\title{Chiral spin chain interfaces exhibiting event horizon physics}

\author{Matthew D. Horner}
\affiliation{School of Physics and Astronomy, University of Leeds, Leeds, LS2 9JT, United Kingdom}

\author{Andrew Hallam}
\affiliation{School of Physics and Astronomy, University of Leeds, Leeds, LS2 9JT, United Kingdom}

\author{Jiannis K. Pachos}
\affiliation{School of Physics and Astronomy, University of Leeds, Leeds, LS2 9JT, United Kingdom}

\date{\today}

\begin{abstract}
The interface between different quantum phases of matter can give rise to novel physics, such as exotic topological phases or non-unitary conformal field theories. Here we investigate the interface between two spin chains in different chiral phases. Surprisingly, the mean field theory approximation of this interacting composite system is given in terms of Dirac fermions in a curved space-time geometry. In particular, the interface between the two phases represents a black hole horizon. We demonstrate that this representation is faithful both analytically, by employing bosonisation to obtain a Luttinger liquid model, and numerically, by employing Matrix Product State methods. A striking prediction from the black hole equivalence emerges when a quench, at one side of the interface between two opposite chiralities, causes the other side to thermalise with the Hawking temperature for a wide range of parameters and initial conditions.
\end{abstract}

\maketitle



Many quantum lattice models exhibit emergent relativistic physics in their continuum limit. The celebrated example is graphene whose low-energy regime is described by the Dirac equation~\cite{Wallace,graphene}. Other examples admitting relativistic descriptions are Kitaev's honeycomb model~\cite{KITAEV20062,Farjami}, superconductors~\cite{PhysRevB.61.10267,Golan_2018} and the XX model~\cite{LIEB1961407,DePasquale}. The relativistic description of these systems opens up the possibility to simulate curved spacetimes in the laboratory.

Here we modify the 1D spin-$1/2$ XX model with a three-spin chiral interaction, making the system interacting. Such chiral systems exhibit a rich spectra of quantum correlations~\cite{Pachos1} and give rise to skyrmionic configurations~\cite{Tikhonov}. We show these chiral systems are effectively modelled by the Dirac equation on a curved spacetime. This gives the possibility to realise a black hole background.

The emergent black hole is revealed by applying the mean field (MF) approximation. We test the validity of this approximation through a detailed analysis of the phase diagram and a comparison with the full spin model. We model the spin model numerically with matrix product state (MPS) techniques, and analytically through bosonisation. We find MF faithfully predicts a phase transition between a chiral and non-chiral phase. Remarkably, the emergent event horizon aligns precisely with chiral phase interfaces. The inside of the black hole corresponds to a chiral region with a central charge of $c = 2$, whilst the outside corresponds to a non-chiral region with a central charge of $c=1$. 

To verify the emergent black hole, we investigate whether it can reproduce the Hawking effect. Hawking radiation is generated by vacuum fluctuations of quantum fields near the horizon of a black hole, which causes the black hole to evaporate~\cite{Hawking,Page_2005}. Here, we stimulate a Hawking-like effect by quenching the MF system. This causes a wavepacket to tunnel across the horizon and escape into the outer region, with a thermal distribution at the Hawking temperature. This alternative mechanism was originally derived in Ref.~\cite{Wilczek} and is used to simulate Hawking radiation in fermionic lattice models~\cite{Volovik1,Volovik2,Volovik3,Yang,Sabsovich,Maertens,Huang,Hang,Guan,Retzker,Rodriguez,Roldan-Molina,Kosior,Steinhauer_2016,Stone_2013,PhysRevResearch.4.043084}. We demonstrate this for a variety of quenches. As the emergent geometry is generated from the couplings, it is fixed with no black hole evaporation or back-reaction of matter on the geometry.

Our investigation shows horizon physics can model chiral interfaces and accurately predicts the evolution of interacting chiral phases across a phase interface. We envision that geometry provides an elegant formalism to model strongly-interacting systems and their interfaces in higher dimensions.

\begin{figure}[t]
\begin{center}
\includegraphics[scale=1]{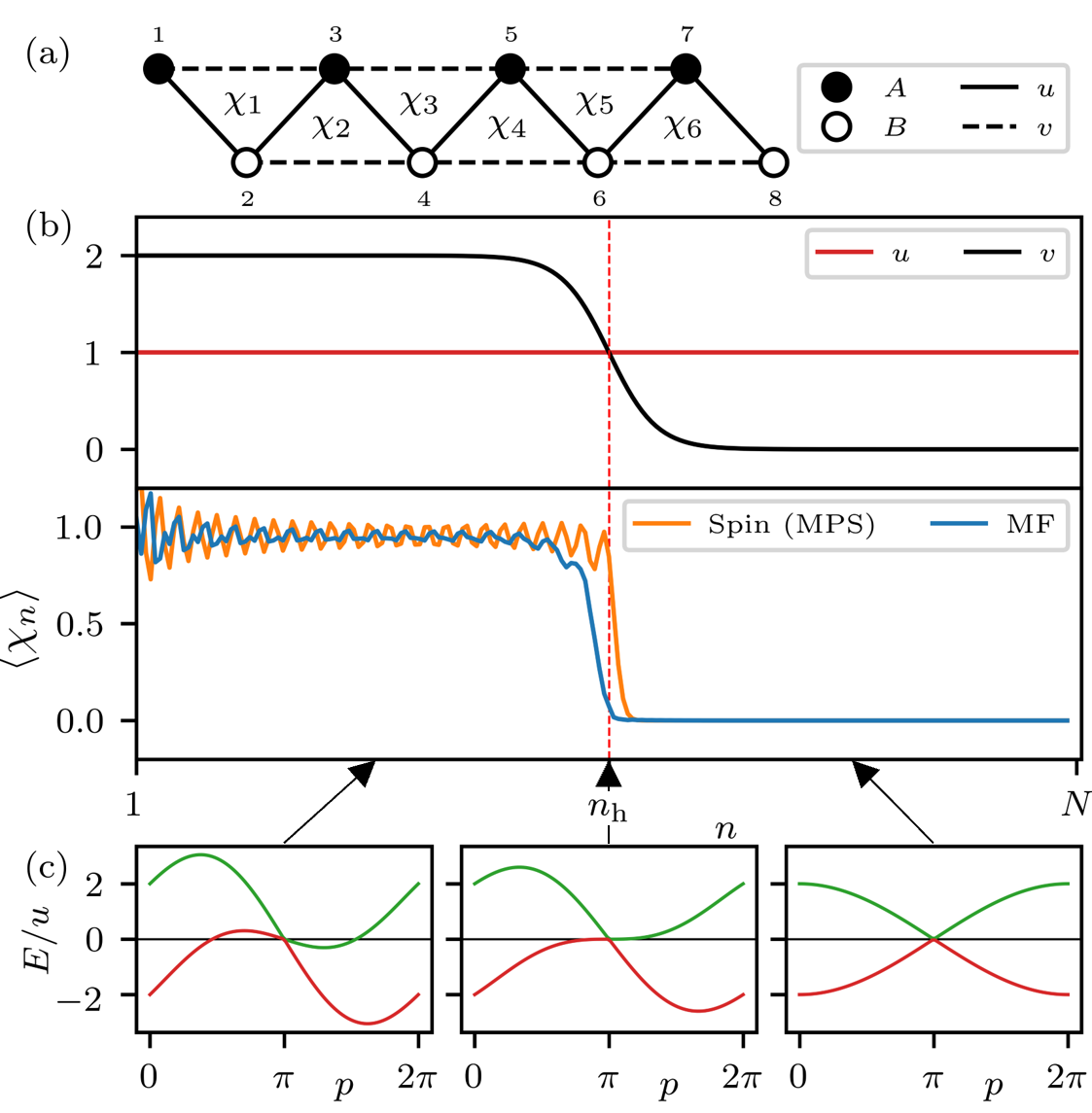}
\end{center}
\caption{(a) The couplings of the Hamiltonian of Eq. \eqref{eq:bh_ham} with diatomic colouring, $A$ and $B$. The chirality operator, $\chi_n$, is defined for each triangular plaquette. (b) An example of interface profile for the couplings $u$ and $v$ with the corresponding ground state chirality, $\langle \chi_n\rangle$. The system changes from a chiral phase, $\langle \chi_n\rangle\neq 0$, when $|v| > |u|$, to non-chiral phase, $\langle \chi_n\rangle=0$, when $|v| < |u|$, across the interface, $n_\mathrm{h}$, corresponding to the event horizon in the continuum theory. The chirality is determined by mean field (MF) theory and by numerical modelling of the spin chain (DMRG with bond dimension $D=400$) for system size $N = 200$. (c) The mean field dispersion relations are obtained from a diatomic representation (see Supplementary Material). The sign of $v$ determines the direction the cones tilt.} 
\label{fig_1} 
\end{figure}

Consider a periodic chain of $N$ spin-$1/2$ particles with Hamiltonian
\begin{equation}
H = \sum_{n=1}^N \left[- \frac{u}{2}  \left(\sigma^x_n \sigma^x_{n+1} + \sigma^y_n \sigma^y_{n+1} \right) + \frac{v}{4} \chi_n \right] 
\label{eq:spin_ham},
\end{equation}
where $u,v \in \mathbb{R}$, $\{ \sigma^x_n , \sigma^y_n , \sigma^z_n \}$ are the Pauli matrices of the $n$th spin and $\chi_n$ is the spin chirality given by the three-spin interaction 
\begin{equation}
\chi_n = \boldsymbol{\sigma}_n \cdot \boldsymbol{\sigma}_{n+1} \times \boldsymbol{\sigma}_{n+2}, 
\end{equation}
where $\boldsymbol{\sigma}_n$ is the vector of Pauli matrices of the $n$th spin \cite{Pachos1,Pachos2}. The chirality operator is a measure of the solid angle spanned by three neighbouring spins. 

The model of Eq.~\eqref{eq:spin_ham} can be mapped to an interacting fermionic Hamiltonian via a Jordan-Wigner transformation. After application of self-consistent mean field (MF) theory to Eq.~\eqref{eq:spin_ham} (see Supplemental Material) we obtain 
\begin{equation}
H_\text{MF} = \sum_{n=1}^N \left( -u c^\dagger_n c_{n+1} - \frac{iv}{2} c_n^\dagger c_{n+2} \right) + \text{h.c.}.
\label{eq:bh_ham}
\end{equation}
This free Hamiltonian is diagonalised exactly with a Fourier transform giving a gapless dispersion with unequal left- and right-moving Fermi velocities. 


By introducing two sublattices, $A$ and $B$, as shown in Fig. \ref{fig_1}(a), the Brillouin zone folds giving two bands as shown in Fig. \ref{fig_1}(c). There is a Dirac cone located at $p_0 = \pi$, which tilts as $v$ is increased. For $|v| < |u|$ we have a type-I Weyl fermion with a single Fermi point; for $|u| = |v|$ we have critical tilting giving rise to a type-III Weyl fermion, with one flat band; and for $|v| > |u|$ the cone over-tilts, corresponding to type-II Weyl fermions~\cite{Volovik1,Volovik2,Volovik3}. For $|v| > |u|$, additional Fermi points appear due to the Nielsen-Ninomiya theorem~\cite{Nielsen1,Nielsen2}. This reveals the emergent relativistic dispersion at low energy.

To reveal the geometric description we derive the corresponding continuum limit. The continuum limit is found by Taylor expanding $H_\text{MF}$ in momentum space about $p_0$, the Fermi point where $E(p_0) = 0$, to first order in $p$~\cite{Farjami,Golan_2018,Tirrito}. This yields a Dirac Hamiltonian on a $(1+1)$D spacetime with metric 
\begin{equation}
\mathrm{d}s^2 =  \left( 1 -\frac{v^2}{u^2} \right) \mathrm{d}t^2 -  \frac{2v}{u^2} \mathrm{d}t \mathrm{d}x - \frac{1}{u^2}\mathrm{d}x^2, \label{eq:metric}
\end{equation}
which is the Gullstrand-Painlev\'{e} metric of a black hole~\cite{Volovik_helium_droplet}. If $v$ is upgraded to a sufficiently slowly-varying function $v(x)$, such as in Fig. \ref{fig_1}(b), then this continuum description remains valid and an event horizon is located at the critical point $x_\mathrm{h}$, where $ |v(x_\mathrm{h})| = |u|$, which corresponds to the location of critical tilt in Fig.~\ref{fig_1}(c). This is a background metric fixed by the couplings, with no back-reaction of the field on the spacetime. From the metric, the corresponding Hawking temperature is~\cite{Volovik3} 
\begin{equation}
T_\text{H} = \frac{1}{2\pi} \left|\frac{\mathrm{d} v(x_\mathrm{h})}{ \mathrm{d}x} \right|. 
\label{eq:Hawking}
\end{equation}
Therefore, the original spin model of Eq.~(\ref{eq:spin_ham}) with inhomogeneous couplings is effectively described by a free fermions on a fixed $(1+1)$D black hole background~\cite{Volovik1,Volovik2,Volovik3,Volovik_helium_droplet}.


\begin{figure}
\begin{center}
\includegraphics[scale=1]{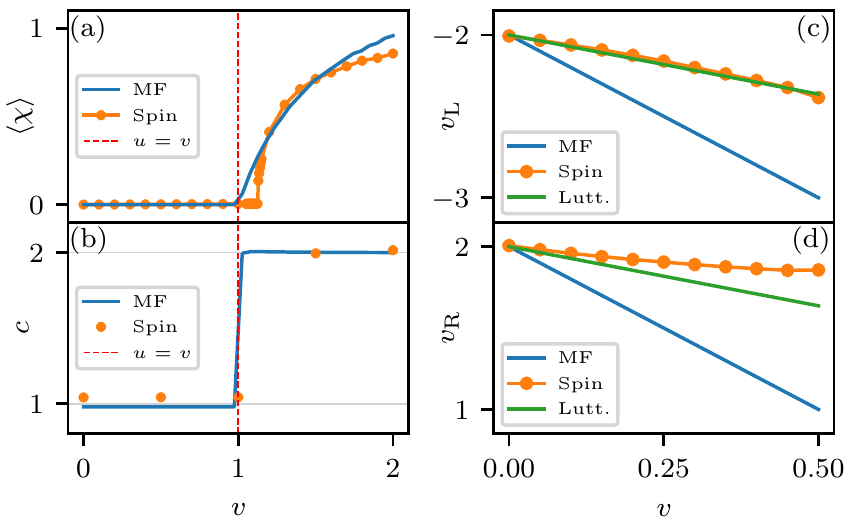}
\end{center}
\caption{(a) The average ground state chirality, $\langle \chi \rangle = \sum_n \langle \chi_n \rangle/N $, for the mean field (MF) model ($N=500$) and spin model found using DMRG ($N=200$, $D=300$) where $u =1$. (b) The central charge, $c$, obtained for the MF model ($N=500$) and the spin model found using DMRG ($v\leq u$: $N=200$, $D=300$ and $v > u$: $N=160$, $D=800$). (c)-(d) The Fermi velocities, $v_\mathrm{L}$ and $v_\mathrm{R}$ respectively ($u=1$) derived from the mean field (MF) and Luttinger liquid descriptions compared to the numerical results of the MPS excitation ansatz for the spin model at bond dimension $D=36$ in the thermodynamic limit.}
\label{fig_2}
\end{figure}

Before proceeding with the black hole analogy, we first establish how accurate the MF approximation is by studying the phase diagram of Eq.~(\ref{eq:spin_ham}) for homogeneous $u$ and $v$. The MF Hamiltonian Eq.~(\ref{eq:bh_ham}) reveals that for $v <u $ the system is in a disordered, gapless, XX phase, while as $v$ increases it passes through a second-order phase transition into a gapless chiral phase, corresponding to a non-zero ground state chirality $\langle \chi_n \rangle$, as Fig.~\ref{fig_2}(a) shows, so chirality is an order parameter. The MF chiral phase transition is located at $|v|=|u|$, coinciding with the critical tilting of the Dirac cones and the appearance of additional Fermi points, as Fig.~\ref{fig_1}(c) shows. Near the critical point, 
\begin{equation}
\langle \chi_n \rangle \sim (v-v_\mathrm{c})^\gamma,
\end{equation}
with critical point $v_c = u$ and critical exponent $\gamma = 1$. Moreover, using finite DMRG \cite{schollwock2011density} we estimate the critical point of the spin model of Eq.~(\ref{eq:spin_ham}) is at $v_\mathrm{c} \approx 1.12u$ with a critical exponent $\gamma \approx 0.39$. A comparison between the chirality of the spin model and the MF model for inhomogeneous and homogeneous couplings can be seen in Fig.~\ref{fig_1}(b) and Fig.~\ref{fig_2}(a) respectively, revealing the effectiveness of MF.

To gain further insight into the chiral phase transition, we consider the behaviour of the entanglement entropy as $v$ is increased. As the model is gapless for all $v$, it can be described by a conformal field theory (CFT). Therefore, the ground state entanglement entropy of a partition of $L \ll N$ spins should follow
\begin{equation}
S_L = \frac{c}{3} \ln L + S_0, \label{eq:entropy}
\end{equation}
where $c \in \mathbb{Z}$ is the central charge of the CFT and $S_0$ is a constant \cite{Calabrese}. Using this formula, we estimate $c$ as a function of $v$ for the full spin model and the MF. In Fig.~\ref{fig_2}(b) we see that $ c \approx 1 $ in the XX phase and $ c \approx 2$ in the chiral phase, with good agreement between the MPS and MF results. We can clearly interpret this in the MF model: the additional Fermi points appearing when $|v| > |u|$, as Fig.~\ref{fig_1}(c) shows, cause the model to transition from a $c=1$ CFT with a single Dirac fermion to a $c=2=1+1$ CFT with two Dirac fermions, as each zero-energy crossing contributes one half of a Dirac fermion. This can also be understood from the lattice structure of the MF model, as Fig.~\ref{fig_1}(a) shows, where for $|v|\ll |u|$ a single zig-zag fermionic chain dominates ($c=1$) while for $|v|\gg |u|$ two fermionic chains dominate, corresponding to the edges of the ladder, thus effectively doubling the degrees of freedom ($c=2$).

The MF faithfully reproduces many features of the full model, especially for $|v| < |u|$ suggesting the interactions are not significant here. We now investigate the validity of the MF analytically by bosonising the spin Hamiltonian for the $|v| < |u|$ phase, and employing Luttinger liquid theory as an alternative derivation of the Fermi velocities of the model~\cite{Giamarchi,Miranda,Sreemayee}. After a Jordan-Wigner transformation, the spin Hamiltonian takes the form $H = H_\mathrm{MF} + H_\mathrm{int}$, where $H_\mathrm{MF}$ is the quadratic MF Hamiltonian of Eq.~(\ref{eq:bh_ham}) and $H_\mathrm{int}$ is an interaction term containing quartic terms. For $|v| < |u|$, the single-band dispersion of the MF Hamiltonian Eq.~(\ref{eq:bh_ham}) suggests the spin model has two Fermi points located at $p_\mathrm{R,L} = \pm \pi/2$, with Fermi velocities $v_\mathrm{R,L} = 2(\pm u - v)$. By expanding around these Fermi points and bosonising the interaction terms using the methods of \cite{Giamarchi,Miranda,Sreemayee}, the fully interacting Hamiltonian is mapped to the free boson Hamiltonian
\begin{equation}
H = u \int \mathrm{d}x \left[ \Pi^2 + (\partial_x \Phi)^2 \right],
\end{equation}
where the fields obey the canonical commutation relations $[\Phi(x), \Pi(y)] = i\delta(x-y)$. The interactions rescale the Fermi velocities $v_\mathrm{R,L} \rightarrow v'_\mathrm{R,L} = 2\left[ \pm u - v\left(1-2/\pi \right) \right]$, but leave the Luttinger parameter unchanged at $K = 1$ (see Supplementary Material), suggesting the model remains a non-interacting free fermion model. 


The dispersion of the spin model as a function of $v$ for $|v| < |u|$ can be calculated using the MPS excitation ansatz working in the thermodynamic limit \cite{haegeman2013post}. This dispersion features unequal left- and right-moving Fermi velocities whose magnitudes change oppositely with $v$ which is the signature of tilting of the cones similar to the MF. In Fig. \ref{fig_2}(c) and \ref{fig_2}(d), the Fermi velocities $v_\mathrm{L,R}$ obtained from MF, the Luttinger liquid model and the spin Hamiltonian are compared. The Luttinger liquid model is more accurate than MF. We expect the disagreement to be lifted at higher order in perturbation theory.

We now study the emergent black hole. It has been shown that many analogue gravitational systems will exhibit a Hawking-like effect, whereby emission of radiation is described by scattering events following a thermal distribution at the Hawking temperature~\cite{Volovik1,Volovik2,Volovik3,Yang,Sabsovich,Maertens,Huang,Hang,Guan,Retzker,Rodriguez,Roldan-Molina,Kosior,Steinhauer_2016,Stone_2013} which is the definition of the Hawking effect we use. Reversing the argument, we investigate whether the Hawking effect can describe quenched time evolutions across the chiral interface. To simulate large system sizes and long evolution times, we resort to the MF of Eq.~(\ref{eq:bh_ham}) rather than the full spin model. Consider an open, inhomogeneous system with couplings $u(x) = 1$ and
\begin{equation}
v(x) = \alpha \tanh[\beta(x-x_\mathrm{h})],
\label{en:couplings}
\end{equation}
where $\alpha ,\beta \in \mathbb{R}$ and $x_\text{h}$ is in the centre of the system. Here, $x$ is the unit cell coordinate in order to align with our continuum conventions (see Supplementary Material). This produces a positive and negative chiral region separated by a small zero-chirality region in between. In the continuum limit, this corresponds to a black hole-white hole interface. This is a common set-up used in the literature, see Refs.~\cite{Roldan-Molina,Volovik_book,Maertens}.

\begin{figure}[t]
\begin{center}
\includegraphics[scale=1]{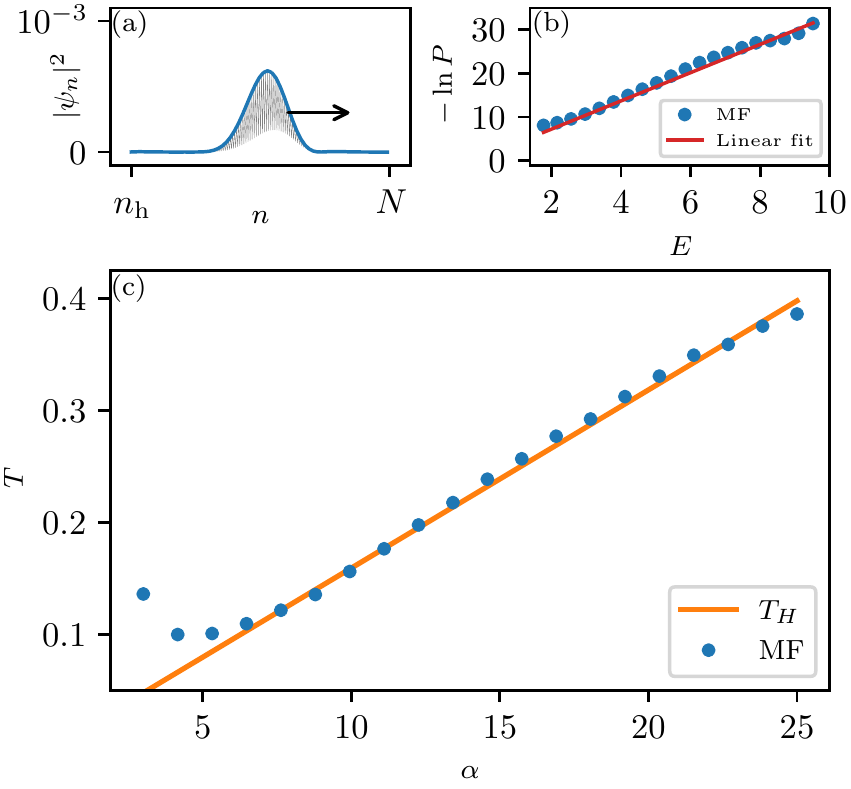}
\end{center}
\caption{(a) The lattice wavefunction $\psi_n$ on the right half of the system ($n \in [n_\text{h},N]$) transmitted through the horizon, for the couplings $u=1$ and $v$ given by Eq. \eqref{en:couplings} with $\alpha = 20$, $\beta = 0.1$, and the horizon at $n_\text{h} = N/2$ with $N=500$. The particle tunnels across at $t \approx 2$ and a wavepacket escapes which we interpret as Hawking radiation. (b) A snapshot of the overlap $-\ln P$ vs. the energy of the state $E$ at time $t = 4.5$. The system thermalises shortly after the particle passes through the interface, displaying a linear dependence on $E$, where the gradient is given by $1/T$. (c) The temperature $T$ of the radiation vs. $\alpha$ extracted after time $t=4.5$. $T$ grows linearly with $\alpha$ close to the predictions of the Hawking formula $T_\text{H} \approx \alpha \beta/2 \pi$.}
\label{fig_3}
\end{figure}

Following the method of Ref. \cite{Yang}, we initialise a single-particle state $|n_0 \rangle = c^\dagger_{n_0}|0\rangle$ on the $n_0$th lattice site inside the left half of the system, and let the wavefunction evolve across the interface into the other half with the Hamiltonian $H_\mathrm{MF}$, as shown in Fig.~\ref{fig_3}(a). We measure the wavefunction overlap with energy modes that exist only on the other side of the interface as
\begin{equation}
P(k,t) = |\langle k | e^{-iHt} |n_0\rangle|^2,
\end{equation}
where $|k \rangle$ are single-particle eigenstates of $H_\mathrm{out}$, where $H_\mathrm{out}$ is the Hamiltonian of Eq.~(\ref{eq:bh_ham}) truncated to the outside region. This method utilises the result that Hawking radiation can be viewed as quantum tunnelling~\cite{Wilczek}, differing from the usual interpretation as vacuum fluctuations. This process will not cause the black hole to evaporate as the effective metric is fixed by the couplings.

We find numerically that interfaces between the two chiral phases thermalises the wavefunction: shortly after the wavefunction evolves across the interface, the external distribution takes the form $P(k,t) \propto e^{-E(k)/T}$, where $T$ is some effective temperature. Fig.~\ref{fig_3}(b) shows $P(k,t)$ at time $t = 4.5$ for a system with parameters $N= 500$, $n_h=250$, $\alpha = 20$ and $\beta = 0.1$, where we prepared the particle at $n_0 = 230$. The value $\beta = 0.1$ is taken to suppress lattice and finite size effects. We see $P(k,t)$ strongly thermalises to a Boltzmann distribution at temperature $T$. In Fig.~\ref{fig_3}(c), we present the dependence of $T$ on $\alpha$. We see it closely follows the Hawking formula $T_\mathrm{H} = \alpha \beta/2 \pi$, obtained from Eq. \eqref{eq:Hawking} and Eq. \eqref{en:couplings}, for a wide range of couplings, $\alpha$, thus accurately modelling the physics of the chiral interface. The thermalisation to $T_\mathrm{H}$ breaks down when $\alpha < 4$ as the couplings are not sharp enough to provide a sufficient interface, whereas for large $\alpha$ the couplings vary too fast for the continuum approximation to be valid, which is where the black hole physics emerges. 

We observe strong thermalisation for the chiral-non-chiral interface, corresponding to a single horizon, such as in Fig.~\ref{fig_1}. However, this system does not thermalise to the Hawking temperature as closely as the black hole-white hole interface as it requires larger system sizes and times than we had numerical access too~\cite{Sabsovich}. Nevertheless, the system only thermalises if it contains a phase interface, or equivalently an event horizon (see Supplemental Material).
\begin{figure}[t!]
\begin{center}
\includegraphics[scale=1]{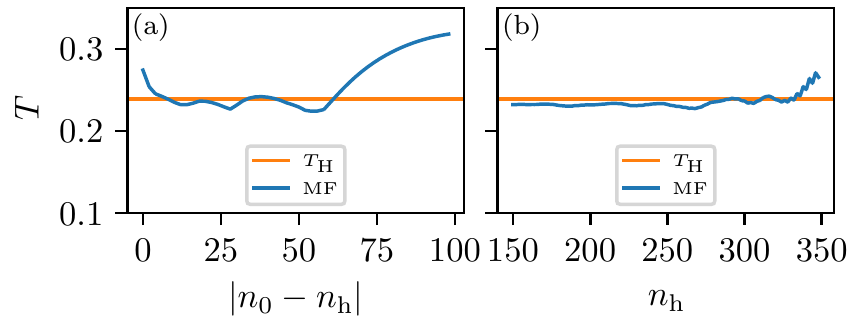}
\end{center}
\caption{(a) The measured temperature $T$ vs. the distance $|n_0 - n_\mathrm{h}|$ from the event horizon that the particle is released for the mean field (MF) system of size $N = 500$, $\alpha = 15$, $\beta = 0.1$ and $n_\mathrm{h} = N/2$. We see the temperature is insensitive to the initial position $n_0$ of the particle. This effect breaks down if $n_0$ is very close to the horizon or too far away, near the boundary of the system. (b) The measured temperature $T$ vs. the position of the horizon $n_\mathrm{h}$ for the same system, where $n_0 = n_\mathrm{h} - 25$.  The temperature is insensitive to the location of the horizon, except when it gets too close to the boundaries, in which case it breaks down.}
\label{fig_4}
\end{figure}

The Hawking temperature $T_\text{H}= \alpha \beta/2 \pi$ is a simple formula that describes a complex thermalisation process. It does not depend on the position of the quench, $n_0$, nor the horizon location, $n_\text{h}$. To verify this numerically, we study the dependence of $T$ on $n_0$ (Fig. \ref{fig_4}(a)) and $n_\text{h}$ (Fig. \ref{fig_4}(b)). We see $T$ is largely insensitive to the initial conditions and only fails if $n_0$ is too close or too far away from the interface, or when the interface $n_\mathrm{h}$ is too close the system edges. In all these cases boundary effects contribute and the exterior region which the overlap $P(k,t)$ is measured in becomes too small. These observations show that the thermalisation is robust, aiding in any potential experimental realisation. We stress this thermalisation is not an equilibration to a thermal state as $t \rightarrow \infty$, but instead is an effective thermalisation due to short time-scale scattering events~\cite{Sabsovich}. As the MF is integrable, it equilibrates to a generalised Gibbs ensemble instead in the $t \rightarrow \infty$ limit~\cite{Perarnau_Llobet,Vidmar,Srednicki2,Deutsch,Rigol} (see Supplemental Material).

In this work, we demonstrated that the low-energy behaviour of a chiral spin chain can be described by Dirac fermions on a black hole background, where the event horizons are aligned with phase interfaces. To demonstrate the faithfulness of this we employed mean field, matrix product states and bosonisation to probe the phase diagram. We simulated the Hawking effect by quenching a system containing a phase interface for a variety of quenches, interface positions and couplings. We envision this bridge between chiral systems and event horizons can facilitate quantum simulations of Hawking radiation, e.g. with cold atom technology \cite{Jaksch,Rodriguez,Kosior}. Moreover, our investigation opens up a way for modelling strongly correlated systems by effective geometric theories with extreme curvature, thus providing a tool for their analytical investigation.

\begin{acknowledgements}
We thank Aydin Deger, Patricio Salgado-Rebolledo, Joe Barker and Diptiman Sen for insightful discussions. M.D.H., A.H. and J.K.P. acknowledge support by EPSRC (Grant No. EP/R020612/1). A.H. acknowledges support by the Leverhulme Trust Research Leadership Award RL-2019-015.
\end{acknowledgements}

\bibliography{refs}

\begin{thebibliography}{48}%
\makeatletter
\providecommand \@ifxundefined [1]{%
 \@ifx{#1\undefined}
}%
\providecommand \@ifnum [1]{%
 \ifnum #1\expandafter \@firstoftwo
 \else \expandafter \@secondoftwo
 \fi
}%
\providecommand \@ifx [1]{%
 \ifx #1\expandafter \@firstoftwo
 \else \expandafter \@secondoftwo
 \fi
}%
\providecommand \natexlab [1]{#1}%
\providecommand \enquote  [1]{``#1''}%
\providecommand \bibnamefont  [1]{#1}%
\providecommand \bibfnamefont [1]{#1}%
\providecommand \citenamefont [1]{#1}%
\providecommand \href@noop [0]{\@secondoftwo}%
\providecommand \href [0]{\begingroup \@sanitize@url \@href}%
\providecommand \@href[1]{\@@startlink{#1}\@@href}%
\providecommand \@@href[1]{\endgroup#1\@@endlink}%
\providecommand \@sanitize@url [0]{\catcode `\\12\catcode `\$12\catcode
  `\&12\catcode `\#12\catcode `\^12\catcode `\_12\catcode `\%12\relax}%
\providecommand \@@startlink[1]{}%
\providecommand \@@endlink[0]{}%
\providecommand \url  [0]{\begingroup\@sanitize@url \@url }%
\providecommand \@url [1]{\endgroup\@href {#1}{\urlprefix }}%
\providecommand \urlprefix  [0]{URL }%
\providecommand \Eprint [0]{\href }%
\providecommand \doibase [0]{https://doi.org/}%
\providecommand \selectlanguage [0]{\@gobble}%
\providecommand \bibinfo  [0]{\@secondoftwo}%
\providecommand \bibfield  [0]{\@secondoftwo}%
\providecommand \translation [1]{[#1]}%
\providecommand \BibitemOpen [0]{}%
\providecommand \bibitemStop [0]{}%
\providecommand \bibitemNoStop [0]{.\EOS\space}%
\providecommand \EOS [0]{\spacefactor3000\relax}%
\providecommand \BibitemShut  [1]{\csname bibitem#1\endcsname}%
\let\auto@bib@innerbib\@empty
\bibitem [{\citenamefont {Wallace}(1947)}]{Wallace}%
  \BibitemOpen
  \bibfield  {author} {\bibinfo {author} {\bibfnamefont {P.~R.}\ \bibnamefont
  {Wallace}},\ }\bibfield  {title} {\bibinfo {title} {The band theory of
  graphite},\ }\href {https://doi.org/10.1103/PhysRev.71.622} {\bibfield
  {journal} {\bibinfo  {journal} {Phys. Rev.}\ }\textbf {\bibinfo {volume}
  {71}},\ \bibinfo {pages} {622} (\bibinfo {year} {1947})}\BibitemShut
  {NoStop}%
\bibitem [{\citenamefont {Castro~Neto}\ \emph {et~al.}(2009)\citenamefont
  {Castro~Neto}, \citenamefont {Guinea}, \citenamefont {Peres}, \citenamefont
  {Novoselov},\ and\ \citenamefont {Geim}}]{graphene}%
  \BibitemOpen
  \bibfield  {author} {\bibinfo {author} {\bibfnamefont {A.~H.}\ \bibnamefont
  {Castro~Neto}}, \bibinfo {author} {\bibfnamefont {F.}~\bibnamefont {Guinea}},
  \bibinfo {author} {\bibfnamefont {N.~M.~R.}\ \bibnamefont {Peres}}, \bibinfo
  {author} {\bibfnamefont {K.~S.}\ \bibnamefont {Novoselov}},\ and\ \bibinfo
  {author} {\bibfnamefont {A.~K.}\ \bibnamefont {Geim}},\ }\bibfield  {title}
  {\bibinfo {title} {The electronic properties of graphene},\ }\href
  {https://doi.org/10.1103/RevModPhys.81.109} {\bibfield  {journal} {\bibinfo
  {journal} {Rev. Mod. Phys.}\ }\textbf {\bibinfo {volume} {81}},\ \bibinfo
  {pages} {109} (\bibinfo {year} {2009})}\BibitemShut {NoStop}%
\bibitem [{\citenamefont {Kitaev}(2006)}]{KITAEV20062}%
  \BibitemOpen
  \bibfield  {author} {\bibinfo {author} {\bibfnamefont {A.}~\bibnamefont
  {Kitaev}},\ }\bibfield  {title} {\bibinfo {title} {Anyons in an exactly
  solved model and beyond},\ }\href
  {https://doi.org/https://doi.org/10.1016/j.aop.2005.10.005} {\bibfield
  {journal} {\bibinfo  {journal} {Annals of Physics}\ }\textbf {\bibinfo
  {volume} {321}},\ \bibinfo {pages} {2} (\bibinfo {year} {2006})},\ \bibinfo
  {note} {january Special Issue}\BibitemShut {NoStop}%
\bibitem [{\citenamefont {Farjami}\ \emph {et~al.}(2020)\citenamefont
  {Farjami}, \citenamefont {Horner}, \citenamefont {Self}, \citenamefont
  {Papi\ifmmode~\acute{c}\else \'{c}\fi{}},\ and\ \citenamefont
  {Pachos}}]{Farjami}%
  \BibitemOpen
  \bibfield  {author} {\bibinfo {author} {\bibfnamefont {A.}~\bibnamefont
  {Farjami}}, \bibinfo {author} {\bibfnamefont {M.~D.}\ \bibnamefont {Horner}},
  \bibinfo {author} {\bibfnamefont {C.~N.}\ \bibnamefont {Self}}, \bibinfo
  {author} {\bibfnamefont {Z.}~\bibnamefont {Papi\ifmmode~\acute{c}\else
  \'{c}\fi{}}},\ and\ \bibinfo {author} {\bibfnamefont {J.~K.}\ \bibnamefont
  {Pachos}},\ }\bibfield  {title} {\bibinfo {title} {Geometric description of
  the kitaev honeycomb lattice model},\ }\href
  {https://doi.org/10.1103/PhysRevB.101.245116} {\bibfield  {journal} {\bibinfo
   {journal} {Phys. Rev. B}\ }\textbf {\bibinfo {volume} {101}},\ \bibinfo
  {pages} {245116} (\bibinfo {year} {2020})}\BibitemShut {NoStop}%
\bibitem [{\citenamefont {Read}\ and\ \citenamefont
  {Green}(2000)}]{PhysRevB.61.10267}%
  \BibitemOpen
  \bibfield  {author} {\bibinfo {author} {\bibfnamefont {N.}~\bibnamefont
  {Read}}\ and\ \bibinfo {author} {\bibfnamefont {D.}~\bibnamefont {Green}},\
  }\bibfield  {title} {\bibinfo {title} {Paired states of fermions in two
  dimensions with breaking of parity and time-reversal symmetries and the
  fractional quantum hall effect},\ }\href
  {https://doi.org/10.1103/PhysRevB.61.10267} {\bibfield  {journal} {\bibinfo
  {journal} {Phys. Rev. B}\ }\textbf {\bibinfo {volume} {61}},\ \bibinfo
  {pages} {10267} (\bibinfo {year} {2000})}\BibitemShut {NoStop}%
\bibitem [{\citenamefont {Golan}\ and\ \citenamefont
  {Stern}(2018)}]{Golan_2018}%
  \BibitemOpen
  \bibfield  {author} {\bibinfo {author} {\bibfnamefont {O.}~\bibnamefont
  {Golan}}\ and\ \bibinfo {author} {\bibfnamefont {A.}~\bibnamefont {Stern}},\
  }\bibfield  {title} {\bibinfo {title} {Probing topological superconductors
  with emergent gravity},\ }\bibfield  {journal} {\bibinfo  {journal} {Physical
  Review B}\ }\textbf {\bibinfo {volume} {98}},\ \href
  {https://doi.org/10.1103/physrevb.98.064503} {10.1103/physrevb.98.064503}
  (\bibinfo {year} {2018})\BibitemShut {NoStop}%
\bibitem [{\citenamefont {Lieb}\ \emph {et~al.}(1961)\citenamefont {Lieb},
  \citenamefont {Schultz},\ and\ \citenamefont {Mattis}}]{LIEB1961407}%
  \BibitemOpen
  \bibfield  {author} {\bibinfo {author} {\bibfnamefont {E.}~\bibnamefont
  {Lieb}}, \bibinfo {author} {\bibfnamefont {T.}~\bibnamefont {Schultz}},\ and\
  \bibinfo {author} {\bibfnamefont {D.}~\bibnamefont {Mattis}},\ }\bibfield
  {title} {\bibinfo {title} {Two soluble models of an antiferromagnetic
  chain},\ }\href
  {https://doi.org/https://doi.org/10.1016/0003-4916(61)90115-4} {\bibfield
  {journal} {\bibinfo  {journal} {Annals of Physics}\ }\textbf {\bibinfo
  {volume} {16}},\ \bibinfo {pages} {407} (\bibinfo {year} {1961})}\BibitemShut
  {NoStop}%
\bibitem [{\citenamefont {De~Pasquale}\ \emph {et~al.}(2008)\citenamefont
  {De~Pasquale}, \citenamefont {Costantini}, \citenamefont {Facchi},
  \citenamefont {Florio}, \citenamefont {Pascazio},\ and\ \citenamefont
  {Yuasa}}]{DePasquale}%
  \BibitemOpen
  \bibfield  {author} {\bibinfo {author} {\bibfnamefont {A.}~\bibnamefont
  {De~Pasquale}}, \bibinfo {author} {\bibfnamefont {G.}~\bibnamefont
  {Costantini}}, \bibinfo {author} {\bibfnamefont {P.}~\bibnamefont {Facchi}},
  \bibinfo {author} {\bibfnamefont {G.}~\bibnamefont {Florio}}, \bibinfo
  {author} {\bibfnamefont {S.}~\bibnamefont {Pascazio}},\ and\ \bibinfo
  {author} {\bibfnamefont {K.}~\bibnamefont {Yuasa}},\ }\bibfield  {title}
  {\bibinfo {title} {Xx model on the circle.},\ }\href
  {https://doi.org/https://doi.org/10.1140/epjst/e2008-00716-9} {\bibfield
  {journal} {\bibinfo  {journal} {Eur. Phys. J. Spec. Top.}\ }\textbf {\bibinfo
  {volume} {160}},\ \bibinfo {pages} {127} (\bibinfo {year}
  {2008})}\BibitemShut {NoStop}%
\bibitem [{\citenamefont {Tsomokos}\ \emph {et~al.}(2008)\citenamefont
  {Tsomokos}, \citenamefont {Garc\'{\i}a-Ripoll}, \citenamefont {Cooper},\ and\
  \citenamefont {Pachos}}]{Pachos1}%
  \BibitemOpen
  \bibfield  {author} {\bibinfo {author} {\bibfnamefont {D.~I.}\ \bibnamefont
  {Tsomokos}}, \bibinfo {author} {\bibfnamefont {J.~J.}\ \bibnamefont
  {Garc\'{\i}a-Ripoll}}, \bibinfo {author} {\bibfnamefont {N.~R.}\ \bibnamefont
  {Cooper}},\ and\ \bibinfo {author} {\bibfnamefont {J.~K.}\ \bibnamefont
  {Pachos}},\ }\bibfield  {title} {\bibinfo {title} {Chiral entanglement in
  triangular lattice models},\ }\href
  {https://doi.org/10.1103/PhysRevA.77.012106} {\bibfield  {journal} {\bibinfo
  {journal} {Phys. Rev. A}\ }\textbf {\bibinfo {volume} {77}},\ \bibinfo
  {pages} {012106} (\bibinfo {year} {2008})}\BibitemShut {NoStop}%
\bibitem [{\citenamefont {Tikhonov}\ \emph {et~al.}(2020)\citenamefont
  {Tikhonov}, \citenamefont {Kondovych}, \citenamefont {Mangeri}, \citenamefont
  {Pavlenko}, \citenamefont {Baudry}, \citenamefont {Sené}, \citenamefont
  {Galda}, \citenamefont {Nakhmanson}, \citenamefont {Heinonen}, \citenamefont
  {Razumnaya}, \citenamefont {Luk’yanchuk},\ and\ \citenamefont
  {Vinokur}}]{Tikhonov}%
  \BibitemOpen
  \bibfield  {author} {\bibinfo {author} {\bibfnamefont {Y.}~\bibnamefont
  {Tikhonov}}, \bibinfo {author} {\bibfnamefont {S.}~\bibnamefont {Kondovych}},
  \bibinfo {author} {\bibfnamefont {J.}~\bibnamefont {Mangeri}}, \bibinfo
  {author} {\bibfnamefont {M.}~\bibnamefont {Pavlenko}}, \bibinfo {author}
  {\bibfnamefont {L.}~\bibnamefont {Baudry}}, \bibinfo {author} {\bibfnamefont
  {A.}~\bibnamefont {Sené}}, \bibinfo {author} {\bibfnamefont
  {A.}~\bibnamefont {Galda}}, \bibinfo {author} {\bibfnamefont
  {S.}~\bibnamefont {Nakhmanson}}, \bibinfo {author} {\bibfnamefont
  {O.}~\bibnamefont {Heinonen}}, \bibinfo {author} {\bibfnamefont
  {A.}~\bibnamefont {Razumnaya}}, \bibinfo {author} {\bibfnamefont
  {I.}~\bibnamefont {Luk’yanchuk}},\ and\ \bibinfo {author} {\bibfnamefont
  {V.~M.}\ \bibnamefont {Vinokur}},\ }\bibfield  {title} {\bibinfo {title}
  {Controllable skyrmion chirality in ferroelectrics},\ }\href
  {https://doi.org/https://doi.org/10.1038/s41598-020-65291-8} {\bibfield
  {journal} {\bibinfo  {journal} {Sci Rep}\ }\textbf {\bibinfo {volume} {10}},\
  \bibinfo {pages} {8657} (\bibinfo {year} {2020})}\BibitemShut {NoStop}%
\bibitem [{\citenamefont {Hawking}(1975)}]{Hawking}%
  \BibitemOpen
  \bibfield  {author} {\bibinfo {author} {\bibfnamefont {S.~W.}\ \bibnamefont
  {Hawking}},\ }\bibfield  {title} {\bibinfo {title} {Particle creation by
  black holes},\ }\href {https://doi.org/https://doi.org/10.1007/BF02345020}
  {\bibfield  {journal} {\bibinfo  {journal} {Communications in Mathematical
  Physics}\ }\textbf {\bibinfo {volume} {43}},\ \bibinfo {pages} {199}
  (\bibinfo {year} {1975})}\BibitemShut {NoStop}%
\bibitem [{\citenamefont {Page}(2005)}]{Page_2005}%
  \BibitemOpen
  \bibfield  {author} {\bibinfo {author} {\bibfnamefont {D.~N.}\ \bibnamefont
  {Page}},\ }\bibfield  {title} {\bibinfo {title} {Hawking radiation and black
  hole thermodynamics*},\ }\href {https://doi.org/10.1088/1367-2630/7/1/203}
  {\bibfield  {journal} {\bibinfo  {journal} {New Journal of Physics}\ }\textbf
  {\bibinfo {volume} {7}},\ \bibinfo {pages} {203} (\bibinfo {year}
  {2005})}\BibitemShut {NoStop}%
\bibitem [{\citenamefont {Parikh}\ and\ \citenamefont
  {Wilczek}(2000)}]{Wilczek}%
  \BibitemOpen
  \bibfield  {author} {\bibinfo {author} {\bibfnamefont {M.~K.}\ \bibnamefont
  {Parikh}}\ and\ \bibinfo {author} {\bibfnamefont {F.}~\bibnamefont
  {Wilczek}},\ }\bibfield  {title} {\bibinfo {title} {Hawking radiation as
  tunneling},\ }\href {https://doi.org/10.1103/PhysRevLett.85.5042} {\bibfield
  {journal} {\bibinfo  {journal} {Phys. Rev. Lett.}\ }\textbf {\bibinfo
  {volume} {85}},\ \bibinfo {pages} {5042} (\bibinfo {year}
  {2000})}\BibitemShut {NoStop}%
\bibitem [{\citenamefont {Volovik}\ and\ \citenamefont
  {Zhang}(2017)}]{Volovik1}%
  \BibitemOpen
  \bibfield  {author} {\bibinfo {author} {\bibfnamefont {G.}~\bibnamefont
  {Volovik}}\ and\ \bibinfo {author} {\bibfnamefont {K.}~\bibnamefont
  {Zhang}},\ }\bibfield  {title} {\bibinfo {title} {Lifshitz transitions,
  type-ii dirac and weyl fermions, event horizon and all that},\ }\href
  {https://doi.org/https://doi.org/10.1007/s10909-017-1817-8} {\bibfield
  {journal} {\bibinfo  {journal} {J. Low Temp. Phys.}\ }\textbf {\bibinfo
  {volume} {189}},\ \bibinfo {pages} {276–299} (\bibinfo {year}
  {2017})}\BibitemShut {NoStop}%
\bibitem [{\citenamefont {Volovik}\ and\ \citenamefont
  {Huhtala}(2002)}]{Volovik2}%
  \BibitemOpen
  \bibfield  {author} {\bibinfo {author} {\bibfnamefont {G.}~\bibnamefont
  {Volovik}}\ and\ \bibinfo {author} {\bibfnamefont {P.}~\bibnamefont
  {Huhtala}},\ }\bibfield  {title} {\bibinfo {title} {Fermionic microstates
  within the painlevé-gullstrand black hole},\ }\href
  {https://doi.org/https://doi.org/10.1134/1.1484981} {\bibfield  {journal}
  {\bibinfo  {journal} {J. Exp. Theor. Phys.}\ }\textbf {\bibinfo {volume}
  {94}},\ \bibinfo {pages} {853–861} (\bibinfo {year} {2002})}\BibitemShut
  {NoStop}%
\bibitem [{\citenamefont {Volovik}(2016)}]{Volovik3}%
  \BibitemOpen
  \bibfield  {author} {\bibinfo {author} {\bibfnamefont {G.}~\bibnamefont
  {Volovik}},\ }\bibfield  {title} {\bibinfo {title} {Black hole and hawking
  radiation by type-ii weyl fermions},\ }\href
  {https://doi.org/https://doi.org/10.1134/S0021364016210050} {\bibfield
  {journal} {\bibinfo  {journal} {Jetp Lett.}\ }\textbf {\bibinfo {volume}
  {104}},\ \bibinfo {pages} {645–648} (\bibinfo {year} {2016})}\BibitemShut
  {NoStop}%
\bibitem [{\citenamefont {Yang}\ \emph {et~al.}(2020)\citenamefont {Yang},
  \citenamefont {Liu}, \citenamefont {Zhu}, \citenamefont {Luo},\ and\
  \citenamefont {Cai}}]{Yang}%
  \BibitemOpen
  \bibfield  {author} {\bibinfo {author} {\bibfnamefont {R.-Q.}\ \bibnamefont
  {Yang}}, \bibinfo {author} {\bibfnamefont {H.}~\bibnamefont {Liu}}, \bibinfo
  {author} {\bibfnamefont {S.}~\bibnamefont {Zhu}}, \bibinfo {author}
  {\bibfnamefont {L.}~\bibnamefont {Luo}},\ and\ \bibinfo {author}
  {\bibfnamefont {R.-G.}\ \bibnamefont {Cai}},\ }\bibfield  {title} {\bibinfo
  {title} {Simulating quantum field theory in curved spacetime with quantum
  many-body systems},\ }\href
  {https://doi.org/10.1103/PhysRevResearch.2.023107} {\bibfield  {journal}
  {\bibinfo  {journal} {Phys. Rev. Research}\ }\textbf {\bibinfo {volume}
  {2}},\ \bibinfo {pages} {023107} (\bibinfo {year} {2020})}\BibitemShut
  {NoStop}%
\bibitem [{\citenamefont {Sabsovich}\ \emph {et~al.}(2022)\citenamefont
  {Sabsovich}, \citenamefont {Wunderlich}, \citenamefont {Fleurov},
  \citenamefont {Pikulin}, \citenamefont {Ilan},\ and\ \citenamefont
  {Meng}}]{Sabsovich}%
  \BibitemOpen
  \bibfield  {author} {\bibinfo {author} {\bibfnamefont {D.}~\bibnamefont
  {Sabsovich}}, \bibinfo {author} {\bibfnamefont {P.}~\bibnamefont
  {Wunderlich}}, \bibinfo {author} {\bibfnamefont {V.}~\bibnamefont {Fleurov}},
  \bibinfo {author} {\bibfnamefont {D.~I.}\ \bibnamefont {Pikulin}}, \bibinfo
  {author} {\bibfnamefont {R.}~\bibnamefont {Ilan}},\ and\ \bibinfo {author}
  {\bibfnamefont {T.}~\bibnamefont {Meng}},\ }\bibfield  {title} {\bibinfo
  {title} {Hawking fragmentation and hawking attenuation in weyl semimetals},\
  }\href {https://doi.org/10.1103/PhysRevResearch.4.013055} {\bibfield
  {journal} {\bibinfo  {journal} {Phys. Rev. Research}\ }\textbf {\bibinfo
  {volume} {4}},\ \bibinfo {pages} {013055} (\bibinfo {year}
  {2022})}\BibitemShut {NoStop}%
\bibitem [{\citenamefont {Maertens}\ \emph {et~al.}(2022)\citenamefont
  {Maertens}, \citenamefont {Bultinck},\ and\ \citenamefont
  {Van~Acoleyen}}]{Maertens}%
  \BibitemOpen
  \bibfield  {author} {\bibinfo {author} {\bibfnamefont {D.}~\bibnamefont
  {Maertens}}, \bibinfo {author} {\bibfnamefont {N.}~\bibnamefont {Bultinck}},\
  and\ \bibinfo {author} {\bibfnamefont {K.}~\bibnamefont {Van~Acoleyen}},\
  }\href {https://doi.org/10.48550/ARXIV.2204.06583} {\bibinfo {title} {Hawking
  radiation on the lattice as universal (floquet) quench dynamics}} (\bibinfo
  {year} {2022})\BibitemShut {NoStop}%
\bibitem [{\citenamefont {Huang}\ \emph {et~al.}(2018)\citenamefont {Huang},
  \citenamefont {Jin},\ and\ \citenamefont {Liu}}]{Huang}%
  \BibitemOpen
  \bibfield  {author} {\bibinfo {author} {\bibfnamefont {H.}~\bibnamefont
  {Huang}}, \bibinfo {author} {\bibfnamefont {K.-H.}\ \bibnamefont {Jin}},\
  and\ \bibinfo {author} {\bibfnamefont {F.}~\bibnamefont {Liu}},\ }\bibfield
  {title} {\bibinfo {title} {Black-hole horizon in the dirac semimetal
  ${\mathrm{zn}}_{2}{\mathrm{in}}_{2}{\mathrm{s}}_{5}$},\ }\href
  {https://doi.org/10.1103/PhysRevB.98.121110} {\bibfield  {journal} {\bibinfo
  {journal} {Phys. Rev. B}\ }\textbf {\bibinfo {volume} {98}},\ \bibinfo
  {pages} {121110} (\bibinfo {year} {2018})}\BibitemShut {NoStop}%
\bibitem [{\citenamefont {Liu}\ \emph {et~al.}(2020)\citenamefont {Liu},
  \citenamefont {Sun}, \citenamefont {Song}, \citenamefont {Huang},
  \citenamefont {Liu},\ and\ \citenamefont {Meng}}]{Hang}%
  \BibitemOpen
  \bibfield  {author} {\bibinfo {author} {\bibfnamefont {H.}~\bibnamefont
  {Liu}}, \bibinfo {author} {\bibfnamefont {J.-T.}\ \bibnamefont {Sun}},
  \bibinfo {author} {\bibfnamefont {C.}~\bibnamefont {Song}}, \bibinfo {author}
  {\bibfnamefont {H.}~\bibnamefont {Huang}}, \bibinfo {author} {\bibfnamefont
  {F.}~\bibnamefont {Liu}},\ and\ \bibinfo {author} {\bibfnamefont
  {S.}~\bibnamefont {Meng}},\ }\bibfield  {title} {\bibinfo {title} {Fermionic
  analogue of high temperature hawking radiation in black phosphorus},\ }\href
  {https://doi.org/10.1088/0256-307X/37/6/067101} {\bibfield  {journal}
  {\bibinfo  {journal} {Chinese Physics Letters}\ }\textbf {\bibinfo {volume}
  {37}},\ \bibinfo {eid} {067101} (\bibinfo {year} {2020})}\BibitemShut
  {NoStop}%
\bibitem [{\citenamefont {Guan}\ \emph {et~al.}(2017)\citenamefont {Guan},
  \citenamefont {Yu}, \citenamefont {Liu}, \citenamefont {Liu}, \citenamefont
  {Dong}, \citenamefont {Lu}, \citenamefont {Yao},\ and\ \citenamefont
  {Yang}}]{Guan}%
  \BibitemOpen
  \bibfield  {author} {\bibinfo {author} {\bibfnamefont {S.}~\bibnamefont
  {Guan}}, \bibinfo {author} {\bibfnamefont {Z.-M.}\ \bibnamefont {Yu}},
  \bibinfo {author} {\bibfnamefont {Y.}~\bibnamefont {Liu}}, \bibinfo {author}
  {\bibfnamefont {G.-B.}\ \bibnamefont {Liu}}, \bibinfo {author} {\bibfnamefont
  {L.}~\bibnamefont {Dong}}, \bibinfo {author} {\bibfnamefont {Y.}~\bibnamefont
  {Lu}}, \bibinfo {author} {\bibfnamefont {Y.}~\bibnamefont {Yao}},\ and\
  \bibinfo {author} {\bibfnamefont {S.~A.}\ \bibnamefont {Yang}},\ }\bibfield
  {title} {\bibinfo {title} {Artificial gravity field, astrophysical analogues,
  and topological phase transitions in strained topological semimetals},\
  }\href {https://doi.org/https://doi.org/10.1038/s41535-017-0026-7} {\bibfield
   {journal} {\bibinfo  {journal} {npj Quantum Materials}\ }\textbf {\bibinfo
  {volume} {2}},\ \bibinfo {pages} {23} (\bibinfo {year} {2017})}\BibitemShut
  {NoStop}%
\bibitem [{\citenamefont {Retzker}\ \emph {et~al.}(2008)\citenamefont
  {Retzker}, \citenamefont {Cirac}, \citenamefont {Plenio},\ and\ \citenamefont
  {Reznik}}]{Retzker}%
  \BibitemOpen
  \bibfield  {author} {\bibinfo {author} {\bibfnamefont {A.}~\bibnamefont
  {Retzker}}, \bibinfo {author} {\bibfnamefont {J.~I.}\ \bibnamefont {Cirac}},
  \bibinfo {author} {\bibfnamefont {M.~B.}\ \bibnamefont {Plenio}},\ and\
  \bibinfo {author} {\bibfnamefont {B.}~\bibnamefont {Reznik}},\ }\bibfield
  {title} {\bibinfo {title} {Methods for detecting acceleration radiation in a
  bose-einstein condensate},\ }\href
  {https://doi.org/10.1103/PhysRevLett.101.110402} {\bibfield  {journal}
  {\bibinfo  {journal} {Phys. Rev. Lett.}\ }\textbf {\bibinfo {volume} {101}},\
  \bibinfo {pages} {110402} (\bibinfo {year} {2008})}\BibitemShut {NoStop}%
\bibitem [{\citenamefont {Rodr\'{\i}guez-Laguna}\ \emph
  {et~al.}(2017)\citenamefont {Rodr\'{\i}guez-Laguna}, \citenamefont
  {Tarruell}, \citenamefont {Lewenstein},\ and\ \citenamefont
  {Celi}}]{Rodriguez}%
  \BibitemOpen
  \bibfield  {author} {\bibinfo {author} {\bibfnamefont {J.}~\bibnamefont
  {Rodr\'{\i}guez-Laguna}}, \bibinfo {author} {\bibfnamefont {L.}~\bibnamefont
  {Tarruell}}, \bibinfo {author} {\bibfnamefont {M.}~\bibnamefont
  {Lewenstein}},\ and\ \bibinfo {author} {\bibfnamefont {A.}~\bibnamefont
  {Celi}},\ }\bibfield  {title} {\bibinfo {title} {Synthetic unruh effect in
  cold atoms},\ }\href {https://doi.org/10.1103/PhysRevA.95.013627} {\bibfield
  {journal} {\bibinfo  {journal} {Phys. Rev. A}\ }\textbf {\bibinfo {volume}
  {95}},\ \bibinfo {pages} {013627} (\bibinfo {year} {2017})}\BibitemShut
  {NoStop}%
\bibitem [{\citenamefont {Rold\'an-Molina}\ \emph {et~al.}(2017)\citenamefont
  {Rold\'an-Molina}, \citenamefont {Nunez},\ and\ \citenamefont
  {Duine}}]{Roldan-Molina}%
  \BibitemOpen
  \bibfield  {author} {\bibinfo {author} {\bibfnamefont {A.}~\bibnamefont
  {Rold\'an-Molina}}, \bibinfo {author} {\bibfnamefont {A.~S.}\ \bibnamefont
  {Nunez}},\ and\ \bibinfo {author} {\bibfnamefont {R.~A.}\ \bibnamefont
  {Duine}},\ }\bibfield  {title} {\bibinfo {title} {Magnonic black holes},\
  }\href {https://doi.org/10.1103/PhysRevLett.118.061301} {\bibfield  {journal}
  {\bibinfo  {journal} {Phys. Rev. Lett.}\ }\textbf {\bibinfo {volume} {118}},\
  \bibinfo {pages} {061301} (\bibinfo {year} {2017})}\BibitemShut {NoStop}%
\bibitem [{\citenamefont {Kosior}\ \emph {et~al.}(2018)\citenamefont {Kosior},
  \citenamefont {Lewenstein},\ and\ \citenamefont {Celi}}]{Kosior}%
  \BibitemOpen
  \bibfield  {author} {\bibinfo {author} {\bibfnamefont {A.}~\bibnamefont
  {Kosior}}, \bibinfo {author} {\bibfnamefont {M.}~\bibnamefont {Lewenstein}},\
  and\ \bibinfo {author} {\bibfnamefont {A.}~\bibnamefont {Celi}},\ }\bibfield
  {title} {\bibinfo {title} {Unruh effect for interacting particles with
  ultracold atoms},\ }\href {https://doi.org/10.21468/SciPostPhys.5.6.061}
  {\bibfield  {journal} {\bibinfo  {journal} {SciPost Phys.}\ }\textbf
  {\bibinfo {volume} {5}},\ \bibinfo {pages} {61} (\bibinfo {year}
  {2018})}\BibitemShut {NoStop}%
\bibitem [{\citenamefont {Steinhauer}(2016)}]{Steinhauer_2016}%
  \BibitemOpen
  \bibfield  {author} {\bibinfo {author} {\bibfnamefont {J.}~\bibnamefont
  {Steinhauer}},\ }\bibfield  {title} {\bibinfo {title} {Observation of quantum
  hawking radiation and its entanglement in an analogue black hole},\ }\href
  {https://doi.org/10.1038/nphys3863} {\ \textbf {\bibinfo {volume} {12}},\
  \bibinfo {pages} {959} (\bibinfo {year} {2016})}\BibitemShut {NoStop}%
\bibitem [{\citenamefont {Stone}(2013)}]{Stone_2013}%
  \BibitemOpen
  \bibfield  {author} {\bibinfo {author} {\bibfnamefont {M.}~\bibnamefont
  {Stone}},\ }\bibfield  {title} {\bibinfo {title} {An analogue of hawking
  radiation in the quantum hall effect},\ }\href
  {https://doi.org/10.1088/0264-9381/30/8/085003} {\bibfield  {journal}
  {\bibinfo  {journal} {Classical and Quantum Gravity}\ }\textbf {\bibinfo
  {volume} {30}},\ \bibinfo {pages} {085003} (\bibinfo {year}
  {2013})}\BibitemShut {NoStop}%
\bibitem [{\citenamefont {Mertens}\ \emph {et~al.}(2022)\citenamefont
  {Mertens}, \citenamefont {Moghaddam}, \citenamefont {Chernyavsky},
  \citenamefont {Morice}, \citenamefont {van~den Brink},\ and\ \citenamefont
  {van Wezel}}]{PhysRevResearch.4.043084}%
  \BibitemOpen
  \bibfield  {author} {\bibinfo {author} {\bibfnamefont {L.}~\bibnamefont
  {Mertens}}, \bibinfo {author} {\bibfnamefont {A.~G.}\ \bibnamefont
  {Moghaddam}}, \bibinfo {author} {\bibfnamefont {D.}~\bibnamefont
  {Chernyavsky}}, \bibinfo {author} {\bibfnamefont {C.}~\bibnamefont {Morice}},
  \bibinfo {author} {\bibfnamefont {J.}~\bibnamefont {van~den Brink}},\ and\
  \bibinfo {author} {\bibfnamefont {J.}~\bibnamefont {van Wezel}},\ }\bibfield
  {title} {\bibinfo {title} {Thermalization by a synthetic horizon},\ }\href
  {https://doi.org/10.1103/PhysRevResearch.4.043084} {\bibfield  {journal}
  {\bibinfo  {journal} {Phys. Rev. Res.}\ }\textbf {\bibinfo {volume} {4}},\
  \bibinfo {pages} {043084} (\bibinfo {year} {2022})}\BibitemShut {NoStop}%
\bibitem [{\citenamefont {D'Cruz}\ and\ \citenamefont
  {Pachos}(2005)}]{Pachos2}%
  \BibitemOpen
  \bibfield  {author} {\bibinfo {author} {\bibfnamefont {C.}~\bibnamefont
  {D'Cruz}}\ and\ \bibinfo {author} {\bibfnamefont {J.~K.}\ \bibnamefont
  {Pachos}},\ }\bibfield  {title} {\bibinfo {title} {Chiral phase from
  three-spin interactions in an optical lattice},\ }\href
  {https://doi.org/10.1103/PhysRevA.72.043608} {\bibfield  {journal} {\bibinfo
  {journal} {Phys. Rev. A}\ }\textbf {\bibinfo {volume} {72}},\ \bibinfo
  {pages} {043608} (\bibinfo {year} {2005})}\BibitemShut {NoStop}%
\bibitem [{\citenamefont {Nielsen}\ and\ \citenamefont
  {Ninomiya}(1981{\natexlab{a}})}]{Nielsen1}%
  \BibitemOpen
  \bibfield  {author} {\bibinfo {author} {\bibfnamefont {H.}~\bibnamefont
  {Nielsen}}\ and\ \bibinfo {author} {\bibfnamefont {M.}~\bibnamefont
  {Ninomiya}},\ }\bibfield  {title} {\bibinfo {title} {A no-go theorem for
  regularizing chiral fermions},\ }\href
  {https://doi.org/https://doi.org/10.1016/0370-2693(81)91026-1} {\bibfield
  {journal} {\bibinfo  {journal} {Physics Letters B}\ }\textbf {\bibinfo
  {volume} {105}},\ \bibinfo {pages} {219} (\bibinfo {year}
  {1981}{\natexlab{a}})}\BibitemShut {NoStop}%
\bibitem [{\citenamefont {Nielsen}\ and\ \citenamefont
  {Ninomiya}(1981{\natexlab{b}})}]{Nielsen2}%
  \BibitemOpen
  \bibfield  {author} {\bibinfo {author} {\bibfnamefont {H.}~\bibnamefont
  {Nielsen}}\ and\ \bibinfo {author} {\bibfnamefont {M.}~\bibnamefont
  {Ninomiya}},\ }\bibfield  {title} {\bibinfo {title} {Absence of neutrinos on
  a lattice: (ii). intuitive topological proof},\ }\href
  {https://doi.org/https://doi.org/10.1016/0550-3213(81)90524-1} {\bibfield
  {journal} {\bibinfo  {journal} {Nuclear Physics B}\ }\textbf {\bibinfo
  {volume} {193}},\ \bibinfo {pages} {173} (\bibinfo {year}
  {1981}{\natexlab{b}})}\BibitemShut {NoStop}%
\bibitem [{\citenamefont {Tirrito}\ \emph {et~al.}(2021)\citenamefont
  {Tirrito}, \citenamefont {Lewenstein},\ and\ \citenamefont
  {Bermudez}}]{Tirrito}%
  \BibitemOpen
  \bibfield  {author} {\bibinfo {author} {\bibfnamefont {E.}~\bibnamefont
  {Tirrito}}, \bibinfo {author} {\bibfnamefont {M.}~\bibnamefont
  {Lewenstein}},\ and\ \bibinfo {author} {\bibfnamefont {A.}~\bibnamefont
  {Bermudez}},\ }\href {https://doi.org/10.48550/ARXIV.2112.07654} {\bibinfo
  {title} {Topological chiral currents in the gross-neveu model extension}}
  (\bibinfo {year} {2021})\BibitemShut {NoStop}%
\bibitem [{\citenamefont {Volovik}(2003)}]{Volovik_helium_droplet}%
  \BibitemOpen
  \bibfield  {author} {\bibinfo {author} {\bibfnamefont {G.~E.}\ \bibnamefont
  {Volovik}},\ }\href@noop {} {\emph {\bibinfo {title} {The Universe in a
  Helium Droplet}}},\ \bibinfo {edition} {2nd}\ ed.\ (\bibinfo  {publisher}
  {Clarendon Press},\ \bibinfo {year} {2003})\ p.\ \bibinfo {pages}
  {424}\BibitemShut {NoStop}%
\bibitem [{\citenamefont {Schollw{\"o}ck}(2011)}]{schollwock2011density}%
  \BibitemOpen
  \bibfield  {author} {\bibinfo {author} {\bibfnamefont {U.}~\bibnamefont
  {Schollw{\"o}ck}},\ }\bibfield  {title} {\bibinfo {title} {The density-matrix
  renormalization group in the age of matrix product states},\ }\href
  {https://doi.org/https://doi.org/10.1016%2Fj.aop.2010.09.012} {\bibfield
  {journal} {\bibinfo  {journal} {Annals of physics}\ }\textbf {\bibinfo
  {volume} {326}},\ \bibinfo {pages} {96} (\bibinfo {year} {2011})}\BibitemShut
  {NoStop}%
\bibitem [{\citenamefont {Calabrese}\ and\ \citenamefont
  {Cardy}(2004)}]{Calabrese}%
  \BibitemOpen
  \bibfield  {author} {\bibinfo {author} {\bibfnamefont {P.}~\bibnamefont
  {Calabrese}}\ and\ \bibinfo {author} {\bibfnamefont {J.}~\bibnamefont
  {Cardy}},\ }\bibfield  {title} {\bibinfo {title} {Entanglement entropy and
  quantum field theory},\ }\href
  {https://doi.org/10.1088/1742-5468/2004/06/p06002} {\bibfield  {journal}
  {\bibinfo  {journal} {Journal of Statistical Mechanics: Theory and
  Experiment}\ }\textbf {\bibinfo {volume} {2004}},\ \bibinfo {pages} {P06002}
  (\bibinfo {year} {2004})}\BibitemShut {NoStop}%
\bibitem [{\citenamefont {Giamarchi}(2003)}]{Giamarchi}%
  \BibitemOpen
  \bibfield  {author} {\bibinfo {author} {\bibfnamefont {T.}~\bibnamefont
  {Giamarchi}},\ }\href
  {https://doi.org/10.1093/acprof:oso/9780198525004.001.0001} {\emph {\bibinfo
  {title} {Quantum Physics in One Dimension}}}\ (\bibinfo  {publisher} {Oxford
  University Press},\ \bibinfo {year} {2003})\BibitemShut {NoStop}%
\bibitem [{\citenamefont {Miranda}(2002)}]{Miranda}%
  \BibitemOpen
  \bibfield  {author} {\bibinfo {author} {\bibfnamefont {E.}~\bibnamefont
  {Miranda}},\ }\bibfield  {title} {\bibinfo {title} {Introduction to
  bosonization},\ }\href@noop {} {\bibfield  {journal} {\bibinfo  {journal}
  {Brazilian Journal of Physics}\ }\textbf {\bibinfo {volume} {33}} (\bibinfo
  {year} {2002})}\BibitemShut {NoStop}%
\bibitem [{\citenamefont {Aditya}\ and\ \citenamefont {Sen}(2021)}]{Sreemayee}%
  \BibitemOpen
  \bibfield  {author} {\bibinfo {author} {\bibfnamefont {S.}~\bibnamefont
  {Aditya}}\ and\ \bibinfo {author} {\bibfnamefont {D.}~\bibnamefont {Sen}},\
  }\bibfield  {title} {\bibinfo {title} {Bosonization study of a generalized
  statistics model with four fermi points},\ }\href
  {https://doi.org/10.1103/PhysRevB.103.235162} {\bibfield  {journal} {\bibinfo
   {journal} {Phys. Rev. B}\ }\textbf {\bibinfo {volume} {103}},\ \bibinfo
  {pages} {235162} (\bibinfo {year} {2021})}\BibitemShut {NoStop}%
\bibitem [{\citenamefont {Haegeman}\ \emph {et~al.}(2013)\citenamefont
  {Haegeman}, \citenamefont {Osborne},\ and\ \citenamefont
  {Verstraete}}]{haegeman2013post}%
  \BibitemOpen
  \bibfield  {author} {\bibinfo {author} {\bibfnamefont {J.}~\bibnamefont
  {Haegeman}}, \bibinfo {author} {\bibfnamefont {T.~J.}\ \bibnamefont
  {Osborne}},\ and\ \bibinfo {author} {\bibfnamefont {F.}~\bibnamefont
  {Verstraete}},\ }\bibfield  {title} {\bibinfo {title} {Post-matrix product
  state methods: To tangent space and beyond},\ }\href
  {https://doi.org/10.1103/PhysRevB.88.075133} {\bibfield  {journal} {\bibinfo
  {journal} {Phys. Rev. B}\ }\textbf {\bibinfo {volume} {88}},\ \bibinfo
  {pages} {075133} (\bibinfo {year} {2013})}\BibitemShut {NoStop}%
\bibitem [{\citenamefont {Volovik}(2009)}]{Volovik_book}%
  \BibitemOpen
  \bibfield  {author} {\bibinfo {author} {\bibfnamefont {G.~E.}\ \bibnamefont
  {Volovik}},\ }\href
  {https://doi.org/10.1093/acprof:oso/9780199564842.001.0001} {\emph {\bibinfo
  {title} {{The Universe in a Helium Droplet}}}}\ (\bibinfo  {publisher}
  {Oxford University Press},\ \bibinfo {year} {2009})\ Chap.~\bibinfo {chapter}
  {32}\BibitemShut {NoStop}%
\bibitem [{\citenamefont {Perarnau-Llobet}\ \emph {et~al.}(2016)\citenamefont
  {Perarnau-Llobet}, \citenamefont {Riera}, \citenamefont {Gallego},
  \citenamefont {Wilming},\ and\ \citenamefont {Eisert}}]{Perarnau_Llobet}%
  \BibitemOpen
  \bibfield  {author} {\bibinfo {author} {\bibfnamefont {M.}~\bibnamefont
  {Perarnau-Llobet}}, \bibinfo {author} {\bibfnamefont {A.}~\bibnamefont
  {Riera}}, \bibinfo {author} {\bibfnamefont {R.}~\bibnamefont {Gallego}},
  \bibinfo {author} {\bibfnamefont {H.}~\bibnamefont {Wilming}},\ and\ \bibinfo
  {author} {\bibfnamefont {J.}~\bibnamefont {Eisert}},\ }\bibfield  {title}
  {\bibinfo {title} {Work and entropy production in generalised gibbs
  ensembles},\ }\href {https://doi.org/10.1088/1367-2630/aa4fa6} {\bibfield
  {journal} {\bibinfo  {journal} {New Journal of Physics}\ }\textbf {\bibinfo
  {volume} {18}},\ \bibinfo {pages} {123035} (\bibinfo {year}
  {2016})}\BibitemShut {NoStop}%
\bibitem [{\citenamefont {Vidmar}\ and\ \citenamefont {Rigol}(2016)}]{Vidmar}%
  \BibitemOpen
  \bibfield  {author} {\bibinfo {author} {\bibfnamefont {L.}~\bibnamefont
  {Vidmar}}\ and\ \bibinfo {author} {\bibfnamefont {M.}~\bibnamefont {Rigol}},\
  }\bibfield  {title} {\bibinfo {title} {Generalized gibbs ensemble in
  integrable lattice models},\ }\href
  {https://doi.org/10.1088/1742-5468/2016/06/064007} {\bibfield  {journal}
  {\bibinfo  {journal} {Journal of Statistical Mechanics: Theory and
  Experiment}\ }\textbf {\bibinfo {volume} {2016}},\ \bibinfo {pages} {064007}
  (\bibinfo {year} {2016})}\BibitemShut {NoStop}%
\bibitem [{\citenamefont {Srednicki}(1994)}]{Srednicki2}%
  \BibitemOpen
  \bibfield  {author} {\bibinfo {author} {\bibfnamefont {M.}~\bibnamefont
  {Srednicki}},\ }\bibfield  {title} {\bibinfo {title} {Chaos and quantum
  thermalization},\ }\href {https://doi.org/10.1103/PhysRevE.50.888} {\bibfield
   {journal} {\bibinfo  {journal} {Phys. Rev. E}\ }\textbf {\bibinfo {volume}
  {50}},\ \bibinfo {pages} {888} (\bibinfo {year} {1994})}\BibitemShut
  {NoStop}%
\bibitem [{\citenamefont {Deutsch}(1991)}]{Deutsch}%
  \BibitemOpen
  \bibfield  {author} {\bibinfo {author} {\bibfnamefont {J.~M.}\ \bibnamefont
  {Deutsch}},\ }\bibfield  {title} {\bibinfo {title} {Quantum statistical
  mechanics in a closed system},\ }\href
  {https://doi.org/10.1103/PhysRevA.43.2046} {\bibfield  {journal} {\bibinfo
  {journal} {Phys. Rev. A}\ }\textbf {\bibinfo {volume} {43}},\ \bibinfo
  {pages} {2046} (\bibinfo {year} {1991})}\BibitemShut {NoStop}%
\bibitem [{\citenamefont {Rigol}\ \emph {et~al.}(2008)\citenamefont {Rigol},
  \citenamefont {Dunjko},\ and\ \citenamefont {Olshanii}}]{Rigol}%
  \BibitemOpen
  \bibfield  {author} {\bibinfo {author} {\bibfnamefont {M.}~\bibnamefont
  {Rigol}}, \bibinfo {author} {\bibfnamefont {V.}~\bibnamefont {Dunjko}},\ and\
  \bibinfo {author} {\bibfnamefont {M.}~\bibnamefont {Olshanii}},\ }\bibfield
  {title} {\bibinfo {title} {Thermalization and its mechanism for generic
  isolated quantum systems},\ }\href {https://doi.org/10.1038/nature06838}
  {\bibfield  {journal} {\bibinfo  {journal} {Nature}\ }\textbf {\bibinfo
  {volume} {452}},\ \bibinfo {pages} {854} (\bibinfo {year}
  {2008})}\BibitemShut {NoStop}%
\bibitem [{\citenamefont {Jaksch}\ \emph {et~al.}(1998)\citenamefont {Jaksch},
  \citenamefont {Bruder}, \citenamefont {Cirac}, \citenamefont {Gardiner},\
  and\ \citenamefont {Zoller}}]{Jaksch}%
  \BibitemOpen
  \bibfield  {author} {\bibinfo {author} {\bibfnamefont {D.}~\bibnamefont
  {Jaksch}}, \bibinfo {author} {\bibfnamefont {C.}~\bibnamefont {Bruder}},
  \bibinfo {author} {\bibfnamefont {J.~I.}\ \bibnamefont {Cirac}}, \bibinfo
  {author} {\bibfnamefont {C.~W.}\ \bibnamefont {Gardiner}},\ and\ \bibinfo
  {author} {\bibfnamefont {P.}~\bibnamefont {Zoller}},\ }\bibfield  {title}
  {\bibinfo {title} {Cold bosonic atoms in optical lattices},\ }\href
  {https://doi.org/10.1103/PhysRevLett.81.3108} {\bibfield  {journal} {\bibinfo
   {journal} {Phys. Rev. Lett.}\ }\textbf {\bibinfo {volume} {81}},\ \bibinfo
  {pages} {3108} (\bibinfo {year} {1998})}\BibitemShut {NoStop}%
\bibitem [{\citenamefont {Nakahara}(2003)}]{Nakahara}%
  \BibitemOpen
  \bibfield  {author} {\bibinfo {author} {\bibfnamefont {M.}~\bibnamefont
  {Nakahara}},\ }\href@noop {} {\emph {\bibinfo {title} {Geometry, Topology and
  Physics}}},\ \bibinfo {edition} {2nd}\ ed.\ (\bibinfo  {publisher} {Taylor
  and Francis},\ \bibinfo {year} {2003})\BibitemShut {NoStop}%
\end{thebibliography}%

\newpage 
\onecolumngrid
\appendix
\section{Mean field theory and its results}
\subsection{Jordan-Wigner transformation}
In this work we study a modification of the 1D spin-$1/2$ XX model. The Hamiltonian is given by
\begin{equation}
H = \frac{1}{2} \sum_{n=1}^N \left[- \frac{u}{2} \left(\sigma^x_n \sigma^x_{n+1} + \sigma^y_n \sigma^y_{n+1} \right) + \frac{v}{4} \chi_n \right] \equiv H_\mathrm{XX} + H_\chi, \label{eq:app_H_full}
\end{equation}
where $u,v \in \mathbb{R}$, $\{ \sigma^x_n , \sigma^y_n , \sigma^z_n \}$ are the Pauli matrices of the $n$th spin and $\chi_n$ is the spin chirality given by the three-spin interaction $\chi_n \equiv \boldsymbol{\sigma}_n \cdot (\boldsymbol{\sigma}_{n+1} \times \boldsymbol{\sigma}_{n+2} )$ \cite{Pachos1,Pachos2}, where $\boldsymbol{\sigma}_n = (\sigma^x_n,\sigma^y_n,\sigma^z_n)$ is the vector of Pauli matrices of the $n$th spin. We apply periodic boundary conditions $\boldsymbol{\sigma}_n = \boldsymbol{\sigma}_{n+N}$ throughout, however we always have the thermodynamic limit $N \rightarrow \infty$ in mind.

In order to make analytic progress with this model, we map from the language of spins to the language of fermions by applying a Jordan-Wigner transformation defined as
\begin{equation}
\sigma_n^+   = \exp\left(-i \pi \sum_{m = 1}^{n-1} c^\dagger_m c_m \right) c^\dagger_n, \quad \sigma_n^-   = \exp\left(i \pi \sum_{m = 1}^{n-1} c^\dagger_m c_m \right) c_n, \quad \sigma^z_n  = 1 - 2c_n^\dagger c_n,
\end{equation}
where $\sigma_n^\pm = (\sigma^x_n \pm i\sigma^y_n)/2$ and $c_n$ are fermionic operators obeying the commutation relations $\{ c_n, c_m^\dagger \} = \delta_{nm}$ and $\{ c_n, c_m \} = \{c_n^\dagger,c_m^\dagger\} = 0$. Using the defintion of $\sigma^\pm_n$, we have the useful identities
\begin{align}
\sigma^x_n \sigma^x_{n+1} + \sigma^y_n \sigma^y_{n+1} & = 2 \sigma^+_n\sigma^-_{n+1} + \mathrm{H.c.}, \\
\sigma^x_n \sigma^x_{n+1} - \sigma^y_n \sigma^y_{n+1} & = 2i \sigma^+_n\sigma^-_{n+1} + \mathrm{H.c.}.
\end{align}
The first identity allows us to rewrite $H_\mathrm{XX}$ straight away, whilst the second identity allows us rewrite the chirality $\chi_n$ as
\begin{equation}
\begin{aligned}
\chi_n &  = \epsilon_{abc} \sigma^a_n \sigma^b_{n+1} \sigma^c_{n+2} \\
& = (\sigma^x_n \sigma^y_{n+1} - \sigma^y_n \sigma^x_{n+1} ) \sigma^z_{n+2} + (\sigma^x_{n+1} \sigma^y_{n+2} - \sigma^y_{n+1} \sigma^x_{n+2} ) \sigma^z_n + (\sigma^x_{n+2} \sigma^y_n  -  \sigma^y_{n+2} \sigma^x_n) \sigma^z_{n+1} \\
& = 2i \left( \sigma^+_n \sigma^-_{n+1} \sigma^z_{n+2} + \sigma^+_{n+1} \sigma^-_{n+2} \sigma^z_n + \sigma^+_{n+2} \sigma^-_n \sigma^z_{n+1}  \right) + \mathrm{H.c.}.
\end{aligned}
\end{equation}
With this, the Hamiltonian of Eq. (\ref{eq:app_H_full}) takes the form
\begin{equation}
H =   \sum_{n=1}^N \left[ -u \sigma^+_n \sigma^-_{n+1} + \frac{iv}{2} \left( \sigma^+_n \sigma^-_{n+1} \sigma^z_{n+2} + \sigma^+_{n+1} \sigma^-_{n+2} \sigma^z_n +  \sigma^+_{n+2} \sigma^-_n \sigma^z_{n+1} \right) \right] + \mathrm{H.c.}.
\end{equation}
Now the Hamiltonian is in a convenient form, we now apply a Jordan-Wigner transformation. We have 
\begin{align}
\sigma^+_n \sigma^-_{n+1} & = c^\dagger_n c_{n+1}, \\
\sigma^+_{n+2} \sigma^-_n & = c^\dagger_{n+2} \exp ( -i \pi c^\dagger_{n+1} c_{n+1} ) c_n, \label{eq:app_JW_2}
\end{align}
therefore the Hamiltonian transforms to
\begin{equation}
H  =  \sum_{n=1}^N \left[ -u c_n^\dagger c_{n+1} +\frac{iv}{2} (c_n^\dagger c_{n+1} \sigma^z_{n+2} + c^\dagger_{n+1} c_{n+2} \sigma^z_n -  c^\dagger_n c_{n+2})\right]  + \text{H.c.}, \label{eq:app_h_interacting}
\end{equation} 
where the final term loses its $\sigma^z_{n+1}$ because $\sigma^z_{n+1} = \exp( i \pi c^\dagger_{n+1}c_{n+1})$ which cancels with the exponential obtained from the Jordan-Wigner transformation in Eq. (\ref{eq:app_JW_2}). We also swap the final term for its Hermitian conjugate which picks up a minus sign. 

For a system with periodic boundary conditions, after applying the Jordan-Wigner transformation, we would pick up boundary terms which couple the last lattice sites $n =N$ and $n = N-1$ to the first lattice sites $n = 1$ and $ n =2$, however these terms contributes an order $O(1/N)$ correction to the Hamiltonian which can we can safely ignore as we assume we work in the thermodynamic limit for large $N$ \cite{DePasquale}.

\subsection{Self-consistency equation}

The Jordan-Wigner transformation brings the Hamiltonian to an interacting fermionic Hamiltonian in Eq. (\ref{eq:app_h_interacting}) due to the four-fermion interaction terms contained in, for example, $c_n^\dagger c_{n+1} \sigma^z_{n+2}$, therefore this Hamiltonian cannot be diagonalised easily. In order to make progress, we apply mean field theory to transform this Hamiltonian into a non-interacting quadratic Hamiltonian. We replace the operators $\sigma^z_n$ with their expectation values as $ \sigma_n^z \rightarrow \langle \sigma^z_n\rangle \equiv Z$, where the expectation value is done with respect to the ground state of the mean field Hamiltonian and we assume translational invariance to drop the index $n$. This gives us the mean field Hamiltonian
\begin{equation}
H_\text{MF}(Z) = \sum_{n=1}^N \left[ -(u - iv Z) c^\dagger_n c_{n+1} - \frac{iv}{2} c^\dagger_n c_{n+2} \right] + \text{H.c.} \equiv \sum_{n,m} h_{nm} c^\dagger_n c_m 
\end{equation}
which is now a quadratic Hamiltonian as a function of $Z$ that can be diagonalised exactly using by diagonalising the single-particle Hamiltonian $h_{ij}$. In order for this to be self-consistent, we require 
\begin{equation}
\langle \Omega(Z)|\sigma^z_n | \Omega(Z)\rangle = Z,\label{eq:self_con}
\end{equation}
where $| \Omega(Z)\rangle$ is the ground state of $H_\text{MF}(Z)$, obtained by occupying all of the negative energy single-particle modes of $h$. Solving this equation numerically, we find that the solution is $Z=0$. This could also be deduced as the original Hamiltonian has particle-hole symmetry which implies half-filling, so for the mean field to respect this property too we take $Z = 0$, see Eq.~(\ref{eq:half_filling}). Hence, our mean-field Hamiltonian is given by
\begin{equation}
H_\text{MF} = \sum_{n=1}^N \left( -u c^\dagger_n c_{n+1} - \frac{iv}{2} c^\dagger_n c_{n+2} \right) + \text{H.c.} \label{eq:app_mf_ham}
\end{equation}

\subsection{Correlations}

\begin{figure}
\begin{center}
\includegraphics[scale=1]{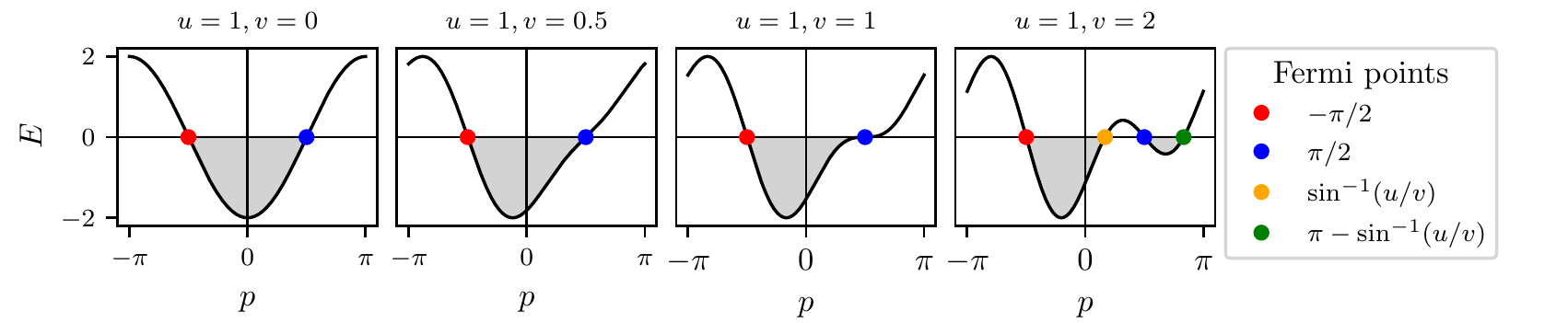}
\end{center}
\caption{The dispersion relation of Eq. (\ref{eq:app_dispersion}) for various values of $v$. We see that two additional Fermi points appear if $v > u$ which divides the negative-energy portion of the Brillouin zone into two disconnected regions.}
\label{fig:appendix_dispersion_1}
\end{figure}
For the case of the homogeneous model (constant $u$ and $v$) the Hamiltonian of Eq. (\ref{eq:app_mf_ham}) has translational symmetry so can be diagonalised exactly with a discrete Fourier transform
\begin{equation}
c_n = \frac{1}{\sqrt{N}} \sum_{p \in \mathrm{B.Z.}} e^{i p n} c_p \label{eq:app_fourier}
\end{equation}
where $\mathrm{B.Z.} = [-\pi,\pi)$ and $k$ is quantised as $k = 2 m \pi/N$ for $m \in \mathbb{Z}$. This yields 
\begin{equation}
H_\mathrm{MF} = \sum_p E(p) c^\dagger_p c_p , \quad E(p) = -2u \cos(p) + v \sin(2p), \label{eq:app_dispersion}
\end{equation} 
where $E(p)$ is the dispersion as shown in Fig. (\ref{fig:appendix_dispersion_1}) for two different values of $v$. The Fermi points of this model are the points for which $E(p) = 0$. We have the usual Fermi points at $p_\mathrm{R,L} = \pm \frac{\pi}{2}$, but if $|v| > |u|$ we have an additional two crossings at $p_1 =  \sin^{-1} \left( \frac{u}{v} \right)$ and $ p_2 = \pi - p_1$. Note that the Fermi velocities at each point $v_\mu = E'(p_\mu)$, where $\mu$ labels the Fermi points, are unequal---the signature of tilting cones as we shall see later. These additional zero energy crossings are a result of the Nielsen-Ninomiya theorem \cite{Nielsen1,Nielsen2} which states that the number of left-movers and right-movers in a lattice model must be equal and they break up the negative energy portion of the Brillouin zone into two disconnected regions as shown by the shaded poprtions in Fig. (\ref{fig:appendix_dispersion_1}). About the two Fermi points $p_\mathrm{R,L}$, the Fermi velocities are given by $v_\mathrm{R,L} = E'(p_\mathrm{R,L}) = 2(\pm u - v)$. 

The correlation matrix is defined as $C_{nm} = \langle \Omega_\mathrm{MF} | c^\dagger_n c_m |\Omega_\mathrm{MF} \rangle$, where $|\Omega_\mathrm{MF} \rangle$ is the ground state of the Hamiltonian of Eq. (\ref{eq:app_mf_ham}). Mapping to momentum space with a discrete Fourier transform of Eq. (\ref{eq:app_fourier}), we can write
\begin{equation}
C_{nm}   = \frac{1}{N} \sum_{p,q \in \text{BZ}} e^{-i p n} e^{i q m} \langle \Omega | c^\dagger_p c_q |\Omega \rangle = \frac{1}{N} \sum_{p:E(p)<0} e^{-ip(n-m)}  \rightarrow \frac{1}{2 \pi} \int_{p:E(p)< 0} \mathrm{d}p e^{-ip(n-m)},
\end{equation} 
where in the second equality we used the fact that the ground state $|\Omega\rangle$ has all negative energy states occupied, so $\langle \Omega | c^\dagger_p c_q |\Omega \rangle = \delta_{pq} \theta(-E(p))$ and in the final equality we took the thermodynamic limit by moulding the sum into a Riemann sum with $\Delta p = 2\pi/N$ and taking the limit as $N \rightarrow \infty$.

In the following calculations we assume that $u,v > 0$. For $v < u$ the correlation function is given by
\begin{equation}
C_{nm} = \frac{1}{2\pi} \int_{-\pi/2}^{\pi/2} \mathrm{d}p e^{-ip(n-m)} = \frac{\sin\left[  (n-m) \frac{\pi}{2} \right] }{\pi (n-m)}, \label{eq:corr_v<u}
\end{equation}
which is independent of $v$ and is the same result obtained for the XX model ($v = 0$). For $v > u$, the negative energy portion of the Brillouin zone splits into two disconnected regions as shown in Fig. \ref{fig:appendix_dispersion_1} so the integral splits into two as
\begin{equation}
C_{nm} =\frac{1}{2\pi} \left( \int_{-\pi/2}^{p_1} \mathrm{d}p + \int_{\pi/2}^{\pi - p_1} \mathrm{d}p \right) e^{-ip(n-m)}  = \frac{i}{2 \pi (n-m)} \left\{ (-1)^{n-m} e^{i p_1 (n-m)} + e^{-ip_1(n-m)} -2 \cos\left[ (n-m)\frac{\pi}{2} \right]  \right\} \label{eq:corr_v>u}
\end{equation}
which is now a function of $v$ and is complex in general. This is the source of the phase transitions to be seen in the next section: due to the change in topology of the negative-energy portion of the Brillouin zone, the correlations are not smooth functions of $v$.
\subsection{Phase transitions}

\begin{figure}
\begin{center}
\includegraphics[scale=0.75]{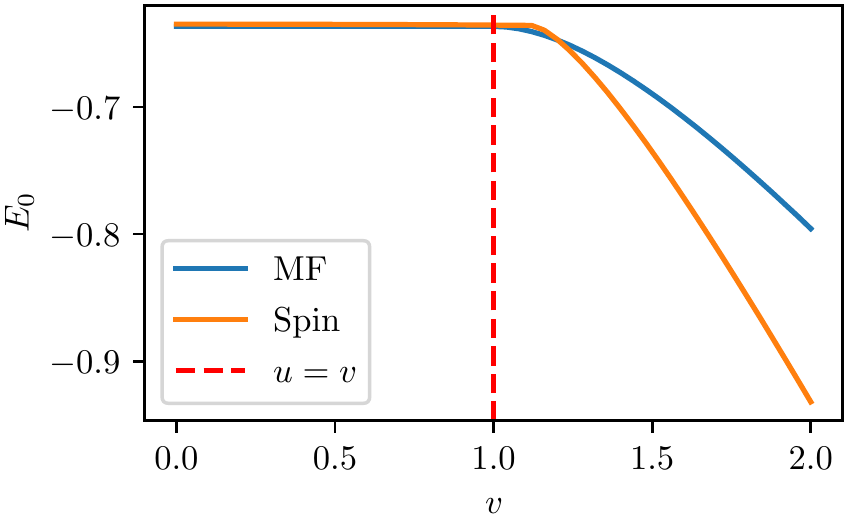}
\caption{A comparison of the ground state energy density vs. $v$ for a sytem of size $N = 200$ and $u = 1$ obtained using the mean field (MF) approximation and the spin model using MPS. This quick test demonstrates that there exists a second-order phase transition about the point $v = u$ and shows that for $v < u$, the effect of the interactions on the model is negligible.}
\label{fig:app_E_vs_v}
\end{center}
\end{figure}

First, we look at the ground state energy density. The density in the thermodynamic limit is given by
\begin{equation}
\rho_0  = \lim_{N \rightarrow \infty} \frac{1}{N} \sum_{p:E(p) < 0} E(p) = \frac{1}{2 \pi} \int_{p:E(p)<0} \mathrm{d}p  E(p) 
\end{equation}
where we took the thermodynamic limit by using the standard trick of moulding the sum into a Riemann sum and taking the limit. Evaluating the integrals for both $v < u$ and $v > u$ yields
\begin{equation}
\rho_0 =  \begin{cases}
  -\frac{2u}{\pi} & v \leq u \\
 -\frac{1}{\pi } \left( \frac{u^2}{v} + v \right) & v > u
\end{cases} .
\end{equation}
We see that $\partial^2 \rho_0/\partial v^2$ is discontinuous at $v = u$ and hence this point corresponds to a second-order phase transition. A comparison of the mean field approximation and spin model obtained through MPS in Fig. \ref{fig:app_E_vs_v} confirms this prediction obtained via the mean field approximation.

Now we look at the chirality. From the mean field description, we found the mean field expression for the chirality operator is given by
\begin{equation}
\chi_n = -2i c^\dagger_n c_{n+2} + \mathrm{H.c.}.
\end{equation}
 Using the correlation matrix Eq. (\ref{eq:corr_v>u}), we find that the chirality is given by
\begin{equation}
\langle \chi_n\rangle  = 4 \mathrm{Im}(C_{n,n+2}) = 
\begin{cases}
0 & v \leq u \\
\frac{4}{\pi} \left( 1 - \frac{u^2}{v^2} \right) & v \geq u
\end{cases},
\end{equation}
so we see that the chirality behaves as an order parameter. Close to the critical point $v = u$ we have
\begin{equation}
\chi_n(v) \approx \chi_n(u) + (v-u) \chi_n'(u) \propto v - u,
\end{equation}
so the critical exponent is equal to $1$.

We see that in order to have a non-zero chirality, we require a complex correlation matrix. In order to achieve this, we require a Hamiltonian that breaks inversion symmetry and has complex next-to-nearest-neighbour correlations, the simplest of which is our Hamiltonian. Of course, this analysis is done using the mean field approximation for the chiral operator, however if we studied the total chirality operator (which contains quartic terms after a Jordan-Wigner transformation) we would have to resort to Wick's theorem in order to evaluate it for a non-interacting system. What we would conclude still is that we require complex next-to-nearest neighbour correlators.

\section{Luttinger model}

\subsection{Particle-hole symmetry}

Let us return to the full spin model of Eq. (\ref{eq:app_H_full}). After a Jordan-Wigner transformation, we arrived at the interacting Hamiltonian
\begin{equation}
H  = \sum_n \left( - u  c_n^\dagger c_{n+1} - \frac{iv}{2} c^\dagger_n c_{n+2} \right)  +\frac{iv}{2} \sum_n  (c_n^\dagger c_{n+1} \sigma^z_{n+2} + c^\dagger_{n+1} c_{n+2} \sigma^z_n )  + \text{H.c.}. \equiv H_0 + H_\mathrm{int}, \label{eq:app_H_full_fermions}
\end{equation}
where $\sigma^z_n = 1 - 2c^\dagger_n c_n$. This fully interacting Hamiltonian has particle-hole symmetry under the transformation
\begin{equation}
c_n \rightarrow U^\dagger c_n U = (-1)^n c_n^\dagger , \quad c_n^\dagger \rightarrow U^\dagger c_n^\dagger U = (-1)^n c_n.
\end{equation}
Let us look at the consequences of this symmetry. Exact diagonalisation of the spin Hamiltonian reveals that the ground state is non-degenerate for an even number of lattice sites, so we take an even $N$ to avoid any subtleties due to degeneracy. Additionally, for the purposes of black hole simulation we always need an even $N$ in order to bi-colour the lattice and form a two-component Dirac spinor. Therefore, our ground state will be an eigenstate of $U$ with an eigenvalue of $\pm 1$ as $U^2 = \mathbb{I}$. Suppose we calculated the ground state density, we have
\begin{equation}
\begin{aligned}
\langle \Omega | c^\dagger_n c_n | \Omega \rangle & = \langle \Omega | U^\dagger c^\dagger_n c_n U|\Omega \rangle \\
& = (-1)^{2n} \langle \Omega | c_n c_n^\dagger | \Omega \rangle \\
& = \langle \Omega |(1 - c^\dagger_n c_n) |\Omega \rangle \\
& = 1 - \langle \Omega | c^\dagger_n c_n |\Omega \rangle \\
\Rightarrow \langle \Omega |c^\dagger_n c_n |\Omega \rangle & = \frac{1}{2} \label{eq:half_filling}
\end{aligned}
\end{equation} 
which is our usual half-filling result. Now, for the nearest-neighbour correlations we have
\begin{equation}
\begin{aligned}
\langle \Omega | c^\dagger_n c_{n+1} |\Omega \rangle & = \langle \Omega|U^\dagger c^\dagger_n c_{n+1} U |\Omega \rangle \\
& = (-1)^{2n + 1} \langle \Omega | c_n c^\dagger_{n+1} |\Omega \rangle \\
& = \langle \Omega | c^\dagger_{n+1} c_n |\Omega \rangle \\
& =  \langle \Omega | c^\dagger_n c_{n+1} |\Omega \rangle^*
\end{aligned}
\end{equation}
therefore the nearest-neighbour correlators are real. We use these results in the following calculation.

For a product of two operators, normal ordering amounts to subtracting off the ground state expectation value as $: A : \ = A - \langle \Omega | A | \Omega \rangle$. We can use this to simplify the interaction term of Eq. (\ref{eq:app_H_full_fermions}) which prepares us for bosonisation later. We have
\begin{equation}
c^\dagger_n c_{n+1} = \ : c^\dagger_n c_{n+1} : + \langle \Omega| c^\dagger_n c_{n+1} |\Omega\rangle \equiv \ :c^\dagger_n c_{n+1} : + \alpha,
\end{equation}
where we have defined the correlation $\alpha = \langle \Omega |c^\dagger_n c_{n+1} |\Omega \rangle$. Similarly, we have
\begin{equation}
\sigma^z_n = 1 - 2 c^\dagger_n c_n =  1 - 2( : c^\dagger_n c_n : + \langle \Omega | c^\dagger_n c_n| \Omega \rangle ) = -2 :c^\dagger_n c_n :,
\end{equation}
where we used the half filling result $\langle \Omega|  c_n^\dagger c_n |\Omega \rangle = \frac{1}{2}$. From this, we can substitute this into the interaction Hamiltonian of Eq. (\ref{eq:app_H_full_fermions}) to give
\begin{equation}
\begin{aligned}
H_\mathrm{int} & = -iv \sum_n \left[ \left( :c^\dagger_n c_{n+1}: + \alpha \right) :c^\dagger_{n+2} c_{n+2}: + \left( :c^\dagger_{n+1} c_{n+2} :+ \alpha \right) :c^\dagger_n c_n: \right] + \mathrm{H.c.} \\
& = -iv \sum_n \left( :c^\dagger_n c_{n+1} : : c^\dagger_{n+2} c_{n+2} : + :c^\dagger_{n+1} c_{n+2} : : c^\dagger_n c_n : \right) + \text{H.c.},
\end{aligned}
\end{equation}
where we used the fact that $\alpha$ is real and $:c^\dagger_n c_n:$ is Hermitian to get rid of $\alpha$.

\subsection{Expanding the fields about the Fermi points}
For the phase $|v| < |u|$, the mean field theory agrees extremely well with the total spin model and demonstrates that the additional chirality term interaction is irrelevant for ground state properties, whereby the model behaves as if it is the XX model ($v = 0$). In this phase, the mean field also suggests that the model has two Fermi points at $p_\mathrm{R,L} = \pm \frac{\pi}{2}$. Therefore, we expand our fields as
\begin{equation}
\frac{c_n}{\sqrt{a}} = \sum_{\mu = \mathrm{R,L}} e^{ip_\mu an} \psi_\mu(x_n), \label{eq:expansion}
\end{equation}
where the sum is over the Fermi points, $\psi_\mu(x)$ is a continuous field sampled at discrete lattice sites and we have reinstated the lattice spacing $a$. 

First, we substitute the expansion of Eq. (\ref{eq:expansion}) into $H_0$ of Eq. (\ref{eq:app_H_full_fermions}) to give
\begin{equation}
H_0  = \sum_{\mu,\nu}\sum_n a e^{-i(p_\mu - p_\nu) an} \left[ -u  e^{ip_\nu a} \psi^\dagger_\mu(x_n)\psi_\nu(x_{n+1}) - \frac{iv}{2}  e^{2ip_\nu a} \psi^\dagger_\mu(x_n) \psi_\nu(x_{n+2}) \right] + \text{H.c.}. 
\end{equation}
We now discard any oscillating term in the Hamiltonian as these integrate to zero, so we  requires $p_\mu = p_\nu$ in the first phase. This yields
\begin{equation}
\begin{aligned}
H_0 & = \sum_{\mu} \sum_n a \left[ -u e^{ip_\mu a}  \psi^\dagger_\mu \left( \psi_\mu + a \partial_x \psi_\mu +O(a^2) \right) -\frac{iv}{2} e^{2ip_\mu a}  \psi^\dagger_\mu \left( \psi_\mu + 2a \partial_x \psi_\mu +O(a^2)\right)  \right]  + \text{H.c.} \\
& = -i \sum_{\mu} \sum_na^2 \left( \pm u \psi^\dagger_\mu \partial_x \psi_\mu  
- v \psi^\dagger_\mu \partial_x \psi_\mu \right) +O(a^3) + \text{H.c.}  \\
&  \rightarrow -2i\sum_{\mu} \int  \mathrm{d}x  v_\mu \psi_\mu^\dagger \partial_x \psi_\mu , 
\end{aligned}
\end{equation}
where in the second line $\pm$ corresponds to $\mu = \mathrm{R,L}$. We have renormalised the couplings as $au \rightarrow u$ and $av \rightarrow v$. We have defined $v_\mathrm{R,L} = 2(\pm u - v)$ which are nothing but the Fermi velocities $v_\mu = E'(p_\mu)$ of the mean field dispersion relation of Eq.~(\ref{eq:app_dispersion}) in the phase $v < u$.

We now repeat the procedure for the interaction term $H_\mathrm{int}$ of Eq. (\ref{eq:app_H_full_fermions}). We substitute in the expansion of Eq. (\ref{eq:expansion}) into $H_\mathrm{int}$ to give
\begin{equation}
H_\text{int} = -iv   \sum_{\mu,\nu,\alpha,\beta} \sum_n  e^{-i(p_\mu - p_\nu + p_\alpha - p_\beta)an} \left( e^{i(p_\nu - 2(p_\alpha - p_\beta))a}+ e^{-i(p_\mu - 2p_\nu)a}\right) :\psi^\dagger_\mu \psi_\nu: : \psi^\dagger_\alpha \psi_\beta : +O(a^3) + \text{H.c.}, \label{eq:int}
\end{equation}
where we have expanded all fields to zeroth order in $a$ to ensure the Hamiltonian retains order $a^2$ and renormalised the couplings as $av \rightarrow v$. We discard any term that oscillates which requires $p_\mu - p_\nu + p_\alpha - p_\beta = 2n \pi/a$ for $n \in \mathbb{Z}$. With this we find only four terms survive giving us
\begin{equation}
H_\text{int}  = 2v \int \mathrm{d}x \left( \rho_\mathrm{R}^2 + \rho_\mathrm{R} \rho_\mathrm{L}  - \rho_\mathrm{L} \rho_\mathrm{R} - \rho_\mathrm{L}^2 \right) + \text{H.c.} = 4v  \int \mathrm{d}x  \left( \rho^2_\mathrm{R} - \rho^2_\mathrm{L} \right),
\end{equation}
where we have defined the normal-ordered densities $\rho_{\mathrm{R,L}} = \ :\psi^\dagger_{\mathrm{R,L}} \psi_{\mathrm{R,L}}:$.
\subsection{Bosonising the Hamiltonian}
If we pull everything together, the normal-ordered Hamiltonian is given by
\begin{equation}
:H: \ = \ :H_0 + H_\text{int}: \ = -i\sum_{\mu = \mathrm{R,L}} \int  \mathrm{d}x  \left( v_\mu : \psi_\mu^\dagger \partial_x \psi_\mu :   \pm 4v  : \rho_\mu^2 : \right)
\end{equation}
where the $\pm$ corresponds to R and L respectively. Following Ref. \cite{Miranda}, we map the fermionic fields $\psi_\mu$ to bosonic fields $\phi_\mu$ with the mapping
\begin{equation}
\psi_\mathrm{R,L} = F_\mathrm{R,L} \frac{1}{\sqrt{2\pi \alpha}} e^{\pm i \frac{2 \pi \hat{N}_\mathrm{R,L} }{L}x} e^{-i \sqrt{2\pi} \phi_\mathrm{R,L}}, \quad \rho_\mathrm{R,L} = \frac{\hat{N}_\mathrm{R,L}}{L} \mp \frac{1}{\sqrt{2\pi}} \partial_x \phi_\mathrm{R,L}
\end{equation}
where $\hat{N}_\mathrm{R,L}$ are defined as the normal ordered number operators for the right- and left-moving excitations respectively, $L = Na$ is the system's length, $F_\mathrm{R,L}$ are a pair of Klein factors and $\alpha$ is a cutoff. The bosonic fields obey the commutation relations
\begin{equation}
[ \phi_\mathrm{R,L}(x),\phi_\mathrm{R,L}(y)] = \pm\frac{i}{2} \mathrm{sgn}(x-y),
\end{equation} 
whilst pairs of fields about different Fermi points commute. The fermionic fields and densities obey the useful identities
\begin{equation}
:\psi^\dagger_\mathrm{R,L} \partial_x \psi_\mathrm{R,L}: = \pm \frac{i}{2}  \partial_x \phi_\mathrm{R,L} , \quad \rho_\mathrm{R,L}  = \mp \frac{1}{\sqrt{2\pi}}\partial_x \phi_\mathrm{R,L},
\end{equation}
where we have taken $L \rightarrow \infty$. With this, the Hamiltonian is mapped to
\begin{equation}
\begin{aligned}
:H: \ &  =  \int \mathrm{d}x \left( \frac{1}{2} \left[ |v_\mathrm{R}| : (\partial_x \phi_\mathrm{R})^2 : + |v_\mathrm{L}| :(\partial_x \phi_\mathrm{L})^2 : \right] + \frac{2v}{\pi} \left[ (\partial_x \phi_\mathrm{R})^2 - (\partial_x \phi_\mathrm{L})^2 \right]  \right) \\
& = \frac{1}{2} \int \mathrm{d}x \left( |v'_\mathrm{R}| :(\partial_x \phi_\mathrm{R})^2 :+ |v_\mathrm{L}'| : (\partial_x \phi_\mathrm{L})^2 : \right), \label{eq:h_boson}
\end{aligned}
\end{equation}
where the renormalised Fermi velocities are given by
\begin{equation}
v_\mathrm{R,L}' = 2 \left[ \pm u - v\left(1 - \frac{2}{\pi}\right) \right].
\end{equation}
As the Fermi velocities of the model are not equal, we must generalise the bosonisation procedure of Ref. \cite{Miranda}. Define the canonical transformation
\begin{equation}
\Phi = \sqrt{\frac{\mathcal{N}}{2}} \left( \sqrt{|v_\mathrm{L}'|} \phi_\mathrm{L} - \sqrt{|v_\mathrm{R}'|} \phi_\mathrm{R} \right), \quad \Theta = \sqrt{\frac{\mathcal{N}}{2}} \left( \sqrt{|v_\mathrm{L}'|}\phi_\mathrm{L} + \sqrt{|v_\mathrm{R}'|}\phi_\mathrm{R} \right), \label{eq:canonical}
\end{equation}
where $\mathcal{N}$ is a constant to ensure the fields obey the correct commutation relations. Just as for the case of equal Fermi velocities in Ref. \cite{Miranda}, we require the fields $\Phi$ and $\Theta$ to obey the commutation relations 
\begin{equation}
[ \Phi(x), \Theta(y) ] = -\frac{i}{2} \mathrm{sgn}(x-y).
\end{equation}
In terms of our canonical transformation, we have
\begin{equation}
\begin{aligned}
[\Phi(x),\Theta(y)] & = \frac{\mathcal{N}}{2} \left( |v'_\mathrm{L}|[\phi_\mathrm{L}(x),\phi_\mathrm{L}(y)] - |v_\mathrm{R}'| [\phi_\mathrm{R}(x),\phi_\mathrm{R}(y)] \right) \\
& = \frac{\mathcal{N}}{2}\left( -\frac{i|v_\mathrm{L}'|}{2} \mathrm{sgn}(x-y) - \frac{i|v_\mathrm{R}'|}{2} \mathrm{sgn}(x-y) \right) \\
& = - \frac{i\mathcal{N}}{4} (|v_\mathrm{L}'| + |v_\mathrm{R}'|) \mathrm{sgn}(x-y),
\end{aligned}
\end{equation}
therefore we require
\begin{equation}
\mathcal{N} = \frac{2}{|v'_\mathrm{L}| + |v_\mathrm{R}'|} = \frac{1}{2u}.
\end{equation}
Inverting the canonical transformation of Eq. (\ref{eq:canonical}), we have
\begin{equation}
\sqrt{|v_\mathrm{L}'|}\phi_- = \sqrt{u} \left( \Theta + \Phi \right) , \quad \sqrt{|v_\mathrm{R}'|}\phi_+ = \sqrt{u} \left( \Theta - \Phi \right).
\end{equation}
Substituting this back into the bosonised Hamiltonian of Eq. (\ref{eq:h_boson}), we have
\begin{equation}
:H: \ = u \int \mathrm{d}x \left[ :(\partial_x \Theta)^2: + :(\partial_x \Phi)^2: \right].
\end{equation}
If we differentiate the commutator $[ \Phi(x),\Theta(y)]$ with respect to $y$, we find $[\Phi(x), \partial_y \Theta(y)] = i \delta(x-y)$, so we can identify the canonical momentum as $\Pi(x) = \partial_x \Theta(x)$. Therefore, the Hamiltonian takes the form of the free boson
\begin{equation}
:H: \ = u \int \mathrm{d}x \left[ :\Pi^2: + :(\partial_x \Phi)^2: \right],
\end{equation}
which is exactly the same result obtained from bosonising the XX model ($v = 0$) which demonstrates that the interactions for $|v| < |u|$ are irrelevant in the ground state. According to the theory of Luttinger liquids, this implies that $K = 1$ which is the sign of non-interacting fermions \cite{Giamarchi}.
\section{Emergent Black Hole}

\begin{figure}
\begin{center}
\includegraphics[scale=1]{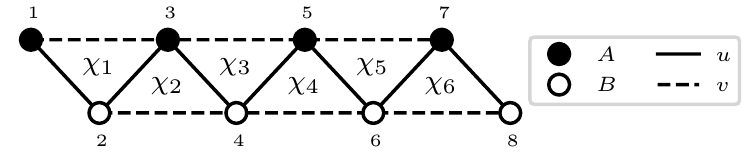}
\caption{In order to see the emergent relativistic description, we introduce a unit cell containing two sites, $A$ and $B$. In this way, the geometry of the lattice can be viewed as a zig-zag ladder, where the rungs join two one-dimensional chains of $A$ and $B$ sites.}
\label{zig_zag_lattice}
\end{center}
\end{figure}
\begin{figure}
\begin{center}
\includegraphics[scale=1]{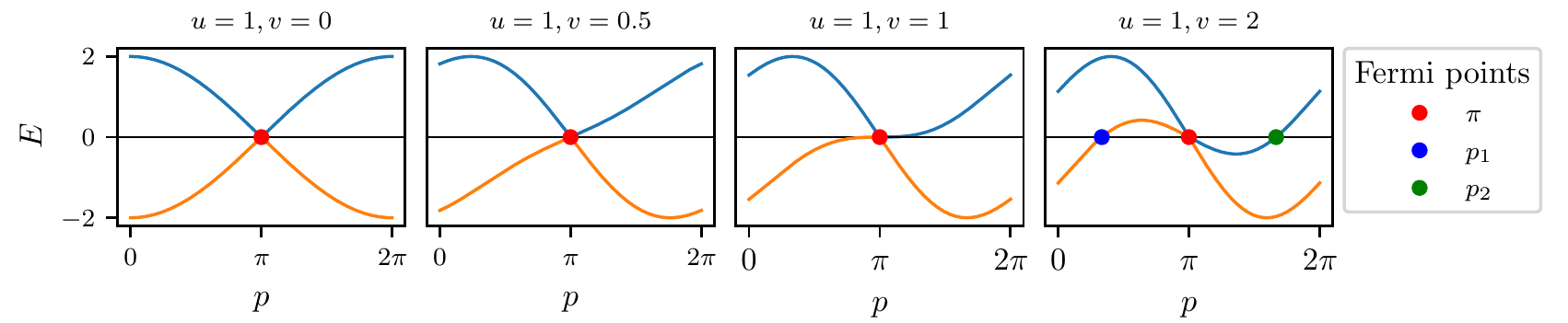}
\end{center}
\caption{The dispersion relation Eq. (\ref{eq:app_dispersion_2}) for various values of $v$ and a fixed $u = 1$. We see that the parameter $v$ tilts the cones in a similar way to a black hole and $v = u$ is the critical value corresponding to the event horizon. Due to the Nielsen-Ninomiya theorem, additional zero-energy crossings appear when the cones over-tilt.} 
\label{fig:appendix_dispersion_2}
\end{figure}

In order to make the link with relativity, we now label the lattice sites as alternating between sub-lattices $A$ and $B$ by introducing a two-site unit cell as shown in Fig.~\ref{zig_zag_lattice}. We can rewrite the mean field Hamiltonian of Eq. (\ref{eq:app_mf_ham}) as 
\begin{equation}
H_\mathrm{MF} =  \sum_n \left[-ua^\dagger_n(b_n + b_{n-1}) - \frac{iv}{2} (a_n^\dagger a_{n+1} + b^\dagger_n b_{n+1}) \right] + \text{H.c.}, \quad u,v \in \mathbb{R} \label{eq:lattice_ham},
\end{equation}
where the Fermions obey the commutation relations $
\{ a_n,a^\dagger_m \}  = \{ b_n, b_m^\dagger\} = \delta_{nm}$, while all other commutators vanish. The index $n$ now labels the unit cells. We Fourier transform the fermions with the definition 
\begin{equation}
a_n = \frac{1}{\sqrt{N_c}} \sum_{p \in \mathrm{B.Z.}} e^{i pa_cn } a_p
\end{equation}
and similarly for $b_n$, where $N_c = N/2$ is the number of unit cells in the system and $a_c = 2a$ is the unit cell spacing for a given lattice spacing $a$. Applying this to the Hamiltonian, we arrive at
\begin{equation}
H_\mathrm{MF} = \sum_p \chi^\dagger_p h(p) \chi_p,
\end{equation}
where we have defined the two-component spinor $\chi_p = (a_p, b_p)^\mathrm{T}$ and the single-particle Hamiltonian
\begin{equation}
h(p) = \begin{pmatrix} g(p)  & f(p) \\ f^*(p) & g(p)  \end{pmatrix}, \quad f(p) = -u(1+e^{-ia_cp}), \quad g(p) = v \sin(a_cp).
\end{equation}
The dispersion relation is given by
\begin{equation}
E(p) = g(p) \pm |f(p)| =  v \sin(a_cp) \pm u \sqrt{2 + 2 \cos(a_c p) } \label{eq:app_dispersion_2}.
\end{equation}
In Fig. \ref{fig:appendix_dispersion_2}, we see that the parameter $v$ has the effect of tilting the cones.

The Fermi points of the dispersion are located at $p_0 = \frac{\pi}{a_c} $ and $p_1 =\frac{1}{a_c} \arccos ( 1 -  \frac{2u^2}{v^2} ) $ however $p_1$ only exists if $|v| \leq |u|$, where the cone is located at $p_1$. Let's focus on the cone: we take the continuum limit by Taylor expanding the single-particle Hamiltonian $h(p)$ about the Fermi point $p_0$. We have
\begin{equation}
h(p_0 + p)  =  a_\mathrm{c} u\sigma^y p - a_\mathrm{c} v \mathbb{I} p + O(p^2) \equiv e_a^{\ i} \alpha^a p_i.
\end{equation}
where we have defined the coefficients $e^x_0 = -a_\mathrm{c} v,e^x_1 = a_\mathrm{c}u $ and the matrices $\alpha^0 = \mathbb{I},\alpha^1 = \sigma^y $. Now we take the continuum limit $a_c \rightarrow 0$ and the thermodynamic limit $N_c \rightarrow \infty$ so both the real space lattice and Brillouin zone become isomorphic to the real line. During the limiting process, we renormalise the couplings $a_cu \rightarrow u$ and $a_c v \rightarrow v$, where $u$ and $v$ are non-zero and finite and we define the continuum limit coordinate $x = na_c$. Note that, due to the bipartite labelling of the lattice, the coordinate $x$ labels the unit cells. Therefore, the continuum limit Hamiltonian after an inverse Fourier transform to real space is given by
\begin{equation}
H  =  \int_\mathbb{R} \mathrm{d}x \chi^\dagger(x) \left( -ie_a^{\ i} \alpha^a \overset{\leftrightarrow}{\partial_i} \right) \chi(x) \label{eq:cont_ham}
\end{equation}
where we have defined $A\overset{\leftrightarrow}{\partial_\mu}B =\frac{1}{2} \left( A \partial_\mu B - (\partial_\mu A)B \right)$ which only acts on spinors, the Dirac alpha and beta matrices $\alpha^a = (\mathbb{I},\sigma^y)$ and $\beta = \sigma^z$. 

The action corresponding to the Hamiltonian of Eq. (\ref{eq:cont_ham}) is given by
\begin{equation}
\begin{aligned}
S & =  \int_M \mathrm{d}^{1+1}x \chi^\dagger(x) \left( i \overset{\leftrightarrow}{\partial_t} + ie_a^{\ i} \alpha^a \overset{\leftrightarrow}{\partial_i}  \right) \chi(x) \\
& =  \int_M \mathrm{d}^{1+1}x  \bar{\chi}(x)  i e^{\ \mu}_a \gamma^a \overset{\leftrightarrow}{\partial_\mu}   \chi(x) ,
\end{aligned} \label{eq:cont_act}
\end{equation}
where we have defined the Dirac gamma matrices $\gamma^0 = \beta = \sigma^z$ and $\gamma^i = \beta \alpha^i$, and $\bar{\chi} = \chi^\dagger \sigma^z$. We see that this action corresponds to the action of a Dirac spinor $\chi$ on a flat Minkowski spacetime with space-dependent parametrs $e_a^{\ \mu}$. 

The quantities $e_a^{\ \mu}$ look very similar to a tetrad basis if we were doing field theory on a curved spacetime. The Dirac action on curved space is given by
\begin{equation}
S =  \int_M \mathrm{d}^{1+1} x |e| \left[ \frac{i}{2} \left( \bar{\psi} \gamma^\mu D_\mu \psi - \overline{D_\mu \psi} \gamma^\mu \psi \right) - m \bar{\psi} \psi \right] ,\label{eq:action1}
\end{equation}
where $D_\mu = \partial_\mu + \omega_\mu$ is the covariant derivative, $\gamma^\mu = e_a^{\ \mu} \gamma^a$ and $\omega_\mu = \frac{1}{8} \omega_{\mu ab} [ \gamma^a, \gamma^b]$, where $e_a^{\ \mu}$ and $\omega_{\mu a b}$ are the components of the the veilbein and spin connection respectively  \cite{Nakahara}. Comparing Eq. (\ref{eq:cont_act}) to Eq. (\ref{eq:action1}), we see that we can interpret the continuum limit of the lattice model as a curved space field theory if we define the spinor $\psi$ related to $\chi$ via
\begin{equation}
\chi = \sqrt{|e|} \psi,
\end{equation}
where $\psi$ is a spinor field which propagates on a spacetime with tetrad
\begin{equation}
e_a^{\ \mu} = \begin{pmatrix} 1 & -v \\ 0 & u \end{pmatrix}, \quad e^a_{\ \mu} = \begin{pmatrix} 1 & v/u \\ 0 & 1/u \end{pmatrix} \label{eq:tetrad}
\end{equation}
and Dirac gamma matrices $\gamma^0  = \sigma^z$ and $\gamma^1 = -i \sigma^x$
which obey the anti-commutation relations $\{ \gamma^a, \gamma^b \} = 2\eta^{ab}$, where $\eta^{ab} = \mathrm{diag}(1,-1)$. The veilbein corresponds to the metric $g^{\mu \nu} = e_a^{\ \mu} e_b^{\ \nu} \eta^{ab}$, which is explicitly given by
\begin{equation}
g^{\mu \nu} = \begin{pmatrix} 1 & -v \\ -v &  v^2    - u^2 \end{pmatrix}, \quad g_{\mu \nu} = \begin{pmatrix}   1 - v^2/u^2  &  -v/u^2 \\ -v/u^2 & -1/u^2 \end{pmatrix},
\end{equation}
or equivalently in terms of differentials
\begin{equation}
\mathrm{d}s^2 =  \left( 1 -\frac{v^2}{u^2} \right) \mathrm{d}t^2 -  \frac{2v}{u^2} \mathrm{d}t \mathrm{d}x - \frac{1}{u^2}\mathrm{d}x^2  \label{eq:app_GP_metric}. 
\end{equation}
This is the Schwarzschild metric expressed in Gullstrand-Painleve coordinates \cite{Volovik_helium_droplet} which is sometimes know as the \textit{acoustic metric}. We refer to this metric as an \textit{internal metric} of the model as it depends upon the internal couplings of the Hamiltonian and not the physical geometry of the lattice.

In order to bring the metric Eq. (\ref{eq:app_GP_metric}) into standard form, we employ the coordinate transformation $(t,x) \mapsto (\tau,x)$ via
\begin{equation}
\tau(t,x)  = t - \int_{x_0}^x \mathrm{d} z  \frac{v(z)}{u^2 - v^2(z)} .
\end{equation}
which maps the metric to
\begin{equation}
\mathrm{d}s^2 =  \left( 1 - \frac{v^2}{u^2} \right) \mathrm{d}\tau^2 -  \frac{1}{u^2 \left( 1 - \frac{v^2}{u^2} \right)} \mathrm{d}x^2,
\end{equation}
which is a metric in Schwarzshild form. If we upgrade $u$ and $v$ to slowly-varying functions of position, then the preceding calculation is still valud and the event horizon is therefore located at the point $x_\mathrm{h}$, where $|v(x_\mathrm{h})| = |u(x_\mathrm{h})|$. In this project, we fix $u(x) = 1$ so it aligns with the standard Schwarzschild metric in natural units. Using the Hawking formula for the temperature of a black hole, the temperature is given by \cite{Volovik3}
\begin{equation}
T_\text{H} = \frac{1}{2\pi} | v'(x_\mathrm{h})|,
\end{equation}
where we have taken $u = 1$.
\section{Thermalisation and Integrability}

In the main text we studied the black hole-white hole interface, corresponding to a chiral and anti-chiral region, with a small non-chiral region between the two. As we see in Figs.~3 and 4 of the text, this thermalises the pulse. This is not the only system that will thermalise however. The requirements on thermalisation is that the system contains a boundary between a chiral and a non-chiral region, in other words, it requires an event horizon. 

\begin{figure}
\begin{center}
\includegraphics[scale=0.9]{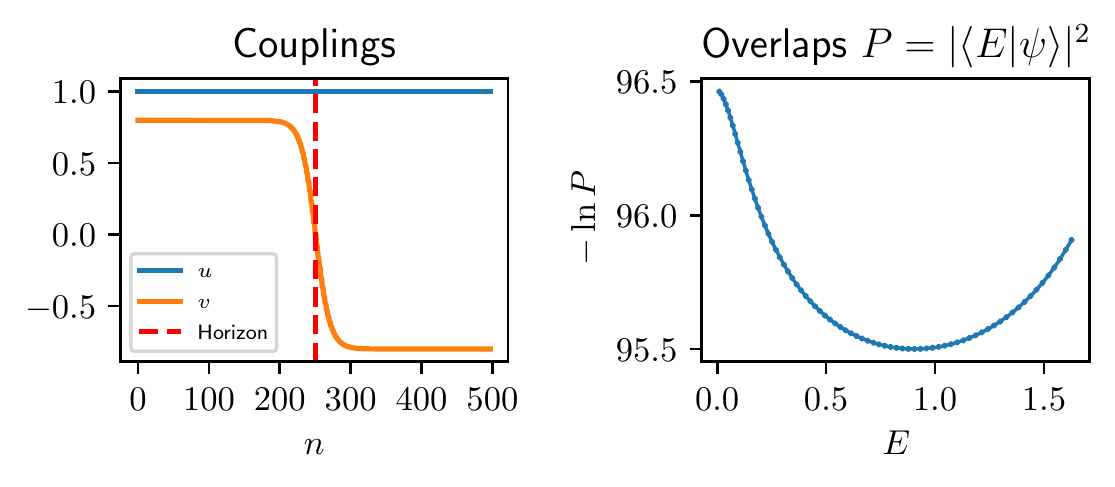}
\includegraphics[scale=0.9]{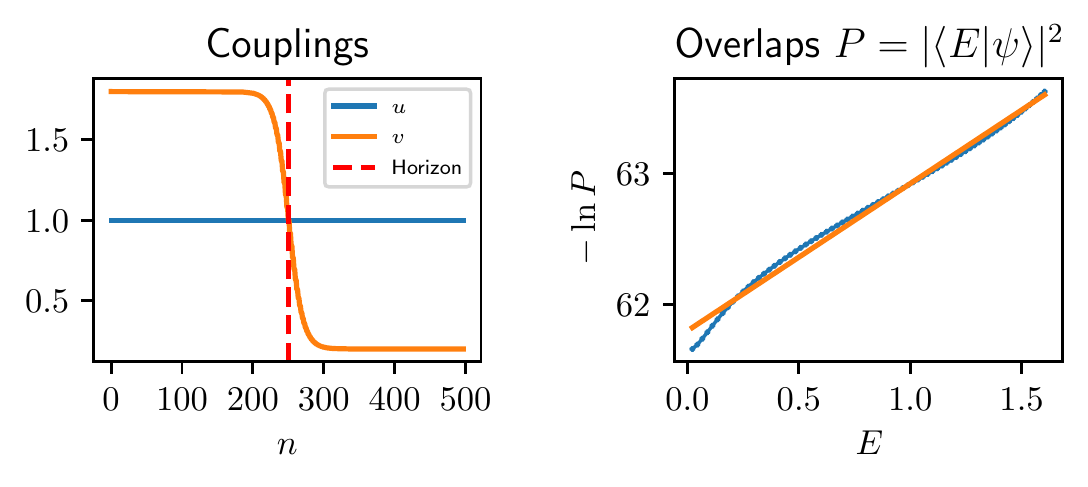}
\includegraphics[scale=0.9]{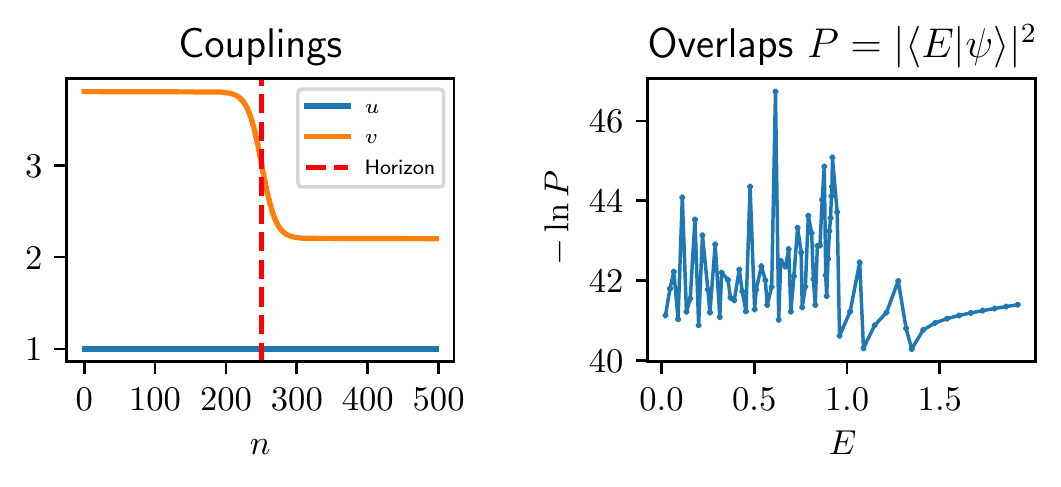}
\end{center}
\caption{The coupling profiles of Eq.~(\ref{couplings}) for $\alpha = 0.8$, $\beta = 0.1$ for $c = 0,1,3$ at the top, middle and bottom respectively. We see that the only system which thermalises strongly is when $c = 1$, which corresponds to a chiral-non-chiral interface, which corresponds to an event horizon. The other two systems do not thermalise as they do not have phase boundaries.}
\label{interfaces}
\end{figure}

We consider couplings of the form $u(x) = 1$ and
\begin{equation}
v(x) = \alpha \tanh[\beta (x - x_\mathrm{h})  ] + c ,\label{couplings}
\end{equation}
where we take $\alpha = 0.8$ and $\beta = 0.1$. The value we take for $c$ determines the type of interface between the two regions. In Fig.~\ref{interfaces} we see the only interface that thermalises is the chiral-non-chiral interface, equivalently the only system that contains an event horizon using the high-energy physics perspective. The chiral-non-chiral interface corresponds to a black hole on the left of the system and the vacuum on the right. This does thermalise strongly, however it does not match the Hawking temperature well as it requires much larger system sizes to achieve this~\cite{Sabsovich}. However, the upshot is that this demonstrates how important the phase boundary/horizon is to the system for thermalisation as systems without a horizon will not thermalise.

It is important to note that, as has been discussed elsewhere, e.g. \cite{Sabsovich}, the Hawking temperature obtained in this study is observed through scattering processes rather than in the equilibration values of observables. The effective thermalisation observed through $H_\mathrm{MF}$ takes place at very short time-scales after release of the particle. If we allow the system to evolve for a long time, it will not equilibrate to a Gibbs ensemble (GE), but instead it will equilibrate to a generalised Gibbs ensemble (GGE) \cite{Perarnau_Llobet,Vidmar}---this is because the mean field Hamiltonian $H_\mathrm{MF}$ is integrable. The generalised GGE is defined as
\begin{equation}
\rho_\text{GGE} = \frac{1}{Z_\text{GGE}}e^{- \sum_k \lambda_k Q_k},\quad Z_\text{GGE} = \mathrm{Tr}\left(e^{- \sum_k \lambda_k Q_k}\right),
\end{equation}
where $\{ Q_k \}$ is a set of conserved charges which commute with the Hamiltonian, $[H,Q_k] = 0$ for all $k$, and $\{ \lambda_k \}$ are their corresponding Lagrange multipliers. The Lagrange multipliers are fixed by ensuring that the charges are conserved, i.e. for an initial pure state $|\psi\rangle$, we require $\langle \psi |Q_k | \psi\rangle = \mathrm{Tr}( \rho_\text{GGE} Q_k )$ which constrains the $\{ \lambda_k \}$. Typically, the charges are given by $Q_k = c^\dagger_k c_k$ which are constructed from the modes which diagonalise the Hamiltonian. 

\begin{figure}[t]
\begin{center}
\includegraphics[scale=1]{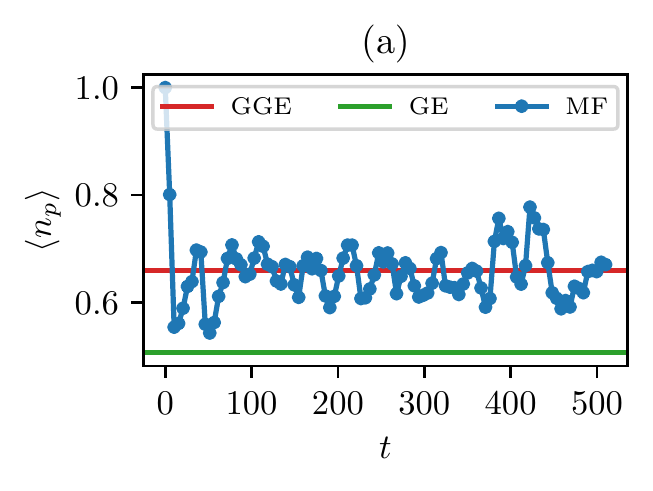}
\includegraphics[scale=1]{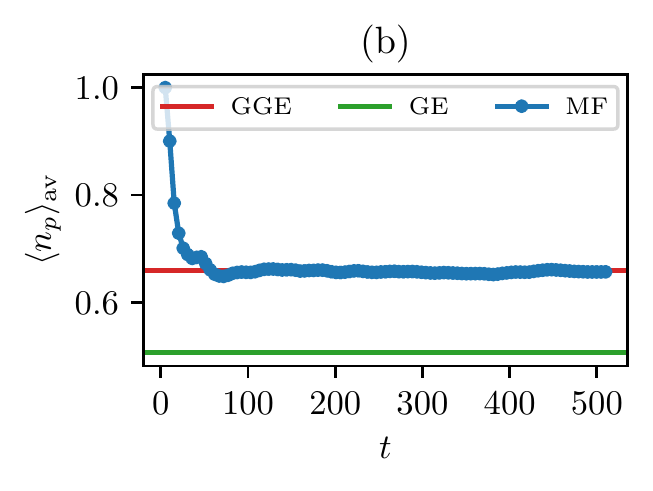}
\end{center}
\caption{(a) The mode occupation of the modes $c_{0,k}$ after quenching from the $v = 0$ ground state (the XX model) for a system of size $N = 200$ for the black hole couplings $v(x)$. We see the system equilibrates to a GGE instead of a GE, therefore we cannot assign a temperature to the system. (b) The time-averaged data $\langle n_p \rangle_\mathrm{av}(t) = \frac{1}{t} \int^t_0 \langle n_p \rangle(t') \mathrm{d}t'$.}
\label{fig:GGE}
\end{figure}

Suppose we looked at the mean field Hamiltonian $H_\mathrm{MF}$. It appears to thermalise the wavefunction of a particle if it passes through the phase boundary between two different chiral regions. It is natural to ask whether this system thermalises in the sense of the eigenstate thermalisation hypothesis~\cite{Srednicki2,Deutsch,Rigol}. If we let the system evolve for a long time, it will equilibrate to a GGE. Consider preparing the system in the ground state $|\Omega_0\rangle$ of the Hamiltonian $H_\text{MF}$ when $v = 0$. In other words, we prepare the system in the ground state of the XX model $H_\mathrm{XX}$ in the absence of any black hole. The initial Hamiltonian is diagonalised to $H_\mathrm{XX} = \sum_k E_0(k)c^\dagger_{0,k}c_{0,k}$. Then we quench by instantaneously switching on the black hole profile $v(x)$ and letting the system evolve with the full Hamiltonian $H_\text{MF}$. We then measure the occupancy of the modes $c_{0,k}$ as time evolves:
\begin{equation}
n_0(k;t) = \langle \Omega_0 | e^{iH_\mathrm{MF}t} c^\dagger_{0,k}c_{0,k} e^{-iH_\mathrm{MF} t} |\Omega_0\rangle.
\end{equation}
In Fig.~\ref{fig:GGE} we find the system equilibrates to the predictions of the GGE and \textit{not} the GE as expected.

An initial analysis of the energy-level statistics of the full spin Hamiltonian of Eq. (\ref{eq:app_H_full}) suggests to us that the model may be integrable---showing Poisson level statistics---however we leave a systematic study of this to future work.

\end{document}